\RequirePackage{fix-cm}
\documentclass[smallextended]{svjour3}       
\smartqed  

\usepackage{bm,amsmath,cite}  
\usepackage{amsfonts} 
\usepackage{graphicx} 

\usepackage[english]{babel}
\usepackage{siunitx,bm}
\usepackage{graphicx,rotating,multirow}
\usepackage{bm,amsmath,amssymb,amsfonts,color,mathrsfs,cite,cuted} 
\usepackage{relsize,exscale}

\journalname{Eur. Phys. J. Plus}
\newcommand{\AJP}{ Am. J. Phys. }
\newcommand{\AnM}{ Annals Math. }
\newcommand{\APB}{ Ann. Phys. (Berlin) }
\newcommand{\APNY}{ Ann. Phys. (New York) }
\newcommand{\CMP}{ Commun. Math. Phys. }
\newcommand{\CPL}{ Chem. Phys. Lett. }
\newcommand{\CRA}{ C. R. Acad. Sci. Ser. A }
\newcommand{\EJP}{ Eur. J. Phys. }
\newcommand{\EPJB}{ Eur. Phys. J. B }
\newcommand{\EPJP}{ Eur. Phys. J. Plus }
\newcommand{\EPJST}{ Eur. Phys. J. Spec. Top. }
\newcommand{\IJQC}{ Int. J. Quantum Chem. }
\newcommand{\JMC}{ J. Math. Chem. }
\newcommand{\JMS}{ J. Mol. Spectrosc. }
\newcommand{\JMP}{ J. Math. Phys. }
\newcommand{\JPA}{ J. Phys. A }
\newcommand{\JSM}{ J. Stat. Mech. }
\newcommand{\JSP}{ J. Stat. Phys. }
\newcommand{\NJP}{ New J. Phys. }
\newcommand{\NL}{ Nature (London) }

\newcommand{\PA}{ Physica A }
\newcommand{\PLA}{ Phys. Lett. A }
\newcommand{\PNAS}{ P. Natl. Acad. Sci. USA }
\newcommand{\PR}{ Phys. Rev. }
\newcommand{\PRA}{ Phys. Rev. A }

\newcommand{\PRD}{ Phys. Rev. D }
\newcommand{\PRL}{ Phys. Rev. Lett. }
\newcommand{\PRE}{ Phys. Rev. E }
\newcommand{\PRe}{ Phys. Rep. }

\newcommand{\PT}{ Phys. Today }

\newcommand{\RMP}{ Rev. Mod. Phys. }
\begin{document}

\title{Quantum information measures of the Dirichlet and Neumann hyperspherical dots\thanks{Research was supported by Competitive Research Project No. 2002143087 from the Research Funding Department, Vice Chancellor for Research and Graduate Studies, University of Sharjah.}
}

\titlerunning{Quantum measures of the Dirichlet and Neumann hyperspherical dots}        

\author{O. Olendski}

\institute{O. Olendski \at
              Department of Applied Physics and Astronomy, University of Sharjah, P.O. Box 27272, Sharjah, United Arab Emirates \\
              \email{oolendski@sharjah.ac.ae}           
}

\date{Received: date / Accepted: date}

\maketitle

\begin{abstract}
$\mathtt{d}$-dimensional hyperspherical quantum dot with either Dirichlet or Neumann boundary conditions (BCs) allows analytic solution of the Schr\"{o}dinger equation in position space and the Fourier transform of the corresponding wave function leads to the analytic form of its momentum counterpart too. This paves the way to an efficient computation in either space of Shannon, R\'{e}nyi and Tsallis entropies, Onicescu energies and Fisher informations; for example, for the latter measure, some particular orbitals exhibit simple expressions in either space at any BC type. A comparative study of the influence of the edge requirement on the quantum information measures proves that the lower threshold of the semi-infinite range of the dimensionless R\'{e}nyi/Tsallis coefficient where one-parameter momentum entropies exist is equal to $\mathtt{d}/(\mathtt{d}+3)$ for the Dirichlet hyperball and $\mathtt{d}/(\mathtt{d}+1)$ for the Neumann one what means that at the unrestricted growth of the dimensionality both measures have their Shannon fellow as the lower verge. Simultaneously, this imposes the restriction on the upper value of the interval $[1/2,\alpha_R)$ inside which the R\'{e}nyi uncertainty relation for the sum of the position $R_\rho(\alpha)$ and wave vector $R_\gamma\left(\frac{\alpha}{2\alpha-1}\right)$ components is defined: $\alpha_R$ is equal to $\mathtt{d}/(\mathtt{d}-3)$ for the Dirichlet geometry and to $\mathtt{d}/(\mathtt{d}-1)$ for the Neumann BC. Some other properties are discussed from mathematical and physical points of view. Parallels are drawn to the corresponding properties of the hydrogen atom and similarities and differences are explained based on the analysis of the associated wave functions.
\keywords{Shannon entropy\and Fisher information\and R\'{e}nyi entropy\and Tsallis entropy\and Quantum dot}
\end{abstract}

\section{Introduction}\label{intro}
Calculation of the Shannon entropies \cite{Shannon1,Shannon2} of the multi-dimensional hydrogen atom and isotropic harmonic oscillator pioneered in the theoretical analysis of quantum-information measures of the structures in position and, accordingly, momentum spaces with more than two coordinates \cite{Yanez1}. Position $S_{\rho_\mathtt{n}}^{(\mathtt{d})}$ and momentum $S_{\gamma_\mathtt{n}}^{(\mathtt{d})}$ Shannon functionals of the $\mathtt{n}$th orbital ($\mathtt{n}=1,2,\ldots$) in the corresponding $\mathtt{d}$-dimensional ($\mathtt{d}$D) spaces are defined as
\begin{subequations}\label{Shannon1}
\begin{align}\label{Shannon1_R}
S_{\rho_\mathtt{n}}^{(\mathtt{d})}&=-\int_{\mathcal{D}_\rho^{(\mathtt{d})}}\rho_\mathtt{n}^{(\mathtt{d})}({\bf r})\ln\rho_\mathtt{n}^{(\mathtt{d})}({\bf r})d{\bf r}\\
\label{Shannon1_K}
S_{\gamma_\mathtt{n}}^{(\mathtt{d})}&=-\int_{\mathcal{D}_\gamma^{(\mathtt{d})}}\gamma_\mathtt{n}^{(\mathtt{d})}({\bf k})\ln\gamma_\mathtt{n}^{(\mathtt{d})}({\bf k})d{\bf k},
\end{align}
\end{subequations} 
with $\mathcal{D}_\rho^{(\mathtt{d})}$ and $\mathcal{D}_\gamma^{(\mathtt{d})}$  being the regions where the position $\Psi_\mathtt{n}^{(\mathtt{d})}({\bf r})$ and wave vector $\Phi_\mathtt{n}^{(\mathtt{d})}({\bf k})$ functions, which are Fourier transforms of each other
\begin{subequations}\label{Fourier1}
\begin{align}\label{Fourier1_1}
\Phi_\mathtt{n}^{(\mathtt{d})}({\bf k})&=\frac{1}{(2\pi)^{\mathtt{d}/2}}\int_{\mathcal{D}_\rho^{(\mathtt{d})}}\Psi_\mathtt{n}^{(\mathtt{d})}({\bf r})e^{-i{\bf kr}}d{\bf r},\\
\label{Fourier1_2}
\Psi_\mathtt{n}^{(\mathtt{d})}({\bf r})&=\frac{1}{(2\pi)^{\mathtt{d}/2}}\int_{\mathcal{D}_\gamma^{(\mathtt{d})}}\Phi_\mathtt{n}^{(\mathtt{d})}({\bf k})e^{i{\bf rk}}d{\bf k},
\end{align}
\end{subequations}
are defined. The waveform $\Psi_\mathtt{n}^{(\mathtt{d})}({\bf r})$ is a solution of the one-particle $\mathtt{d}$D Schr\"{o}dinger equation
\begin{equation}\label{Schrodinger1}
\hat{H}^{(\mathtt{d})}\Psi_\mathtt{n}^{(\mathtt{d})}({\bf r})=E_\mathtt{n}^{(\mathtt{d})}\Psi_\mathtt{n}^{(\mathtt{d})}({\bf r})
\end{equation}
with the Hamiltonian
\begin{equation}\label{Hamiltonian1}
\hat{H}^{(\mathtt{d})}=-\frac{\hbar^2}{2m^*}{\bm\nabla}_{\bf r}^2+V({\bf r}),
\end{equation}
where $V({\bf r})$ is the external potential in which the particle with  mass $m^*$ and energy $E_\mathtt{n}^{(\mathtt{d})}$ is moving and the $\mathtt{d}$D Laplace operator $\Delta\equiv{\bm\nabla}_{\bf r}^2$ in the Cartesian coordinates $(x_1,x_2,\ldots,x_\mathtt{d})$ takes a standard form: $\Delta=\sum_{j=1}^\mathtt{d}\frac{\partial^2}{\partial x_j^2}$. Wave functions $\Psi_\mathtt{n}^{(\mathtt{d})}({\bf r})$ and $\Phi_\mathtt{n}^{(\mathtt{d})}({\bf k})$ are orthonormalized:
\begin{equation}\label{OrthoNormality1}
\int_{\mathcal{D}_\rho^{(\mathtt{d})}}{\Psi_\mathtt{n'}^{(\mathtt{d})}}^\ast({\bf r}){\Psi_\mathtt{n}^{(\mathtt{d})}}d{\bf r}=\int_{\mathcal{D}_\gamma^{(\mathtt{d})}}{\Phi_\mathtt{n'}^{(\mathtt{d})}}^*({\bf k})\Phi_\mathtt{n}^{(\mathtt{d})}({\bf k})d{\bf k}=\delta_\mathtt{n'n},
\end{equation}
with $\delta_\mathtt{n'n}$ being a Kronecker delta, $\mathtt{n'}=1,2,\ldots$,  and they define the corresponding densities $\rho_\mathtt{n}^{(\mathtt{d})}({\bf r})$ and $\gamma_\mathtt{n}^{(\mathtt{d})}({\bf k})$ from Eqs.~\eqref{Shannon1} as $\rho_\mathtt{n}^{(\mathtt{d})}({\bf r})=\left|\Psi_\mathtt{n}^{(\mathtt{d})}({\bf r})\right|^2$ and $\gamma_\mathtt{n}^{(\mathtt{d})}({\bf k})=\left|\Phi_\mathtt{n}^{(\mathtt{d})}({\bf k})\right|^2$. Important property relating position and momentum components is the fact that their sum
\begin{equation}\label{ShannonSum1}
S_t^{(\mathtt{d})}=S_\rho^{(\mathtt{d})}+S_\gamma^{(\mathtt{d})}
\end{equation}
for each orbital can not be smaller than the $\mathtt{d}$-dependent fundamental limit \cite{Bialynicki2,Beckner2}:
\begin{equation}\label{ShannonInequality1}
S_{t_\mathtt{n}}^{(\mathtt{d})}\geq\mathtt{d}(1+\ln\pi).
\end{equation}

If the system possesses some characteristic length $L$, the corresponding entropies are represented as
\begin{subequations}\label{ShannonDimensionless1}
\begin{align}\label{ShannonDimensionless1_rho}
S_{\rho_\mathtt{n}}^{(\mathtt{d})}&=\mathtt{d}\ln L+\overline{S}_{\rho_\mathtt{n}}^{(\mathtt{d})},\\
\label{ShannonDimensionless1_gamma}
S_{\gamma_\mathtt{n}}^{(\mathtt{d})}&=-\mathtt{d}\ln L+\overline{S}_{\gamma_\mathtt{n}}^{(\mathtt{d})},
\end{align}
\end{subequations}
with the overline denoting a dimensionless quantity. Eqs.~\eqref{ShannonDimensionless1} mean that the position and momentum Shannon components are measured in units of the logarithm of the length times dimensionality whereas their sum $S_t^{(\mathtt{d})}$ is a dimensionless scale-independent quantity. Presence of the logarithm makes the Shannon entropy an additive measure:
\begin{equation}\label{Additive1}
S_{fg}=S_f+S_g,
\end{equation}
where $f$ and $g$ are probability functions of the two independent events.

Shannon entropy describes quantitatively the localization/delocalization of the particle in the corresponding domain $\mathcal{D}_\rho^{(\mathtt{d})}$ or $\mathcal{D}_\gamma^{(\mathtt{d})}$: the smaller (larger) its value is, the more (less) information is available about the particle behavior. From this point of view, inequality~\eqref{ShannonInequality1} manifests that the simultaneous knowledge with the arbitrary small precision of both position and momentum ${\bf p}\equiv\hbar{\bf k}$ can not be achieved. Besides this fundamental relation, Shannon entropy is indispensable in many other fields of nano physics; for example, its evaluation on eigenfunctions of quantum systems has been investigated in connection with a method for the approximate description of pure states based on the maximum entropy principle \cite{Plastino1}. For the hydrogen atom, it was shown, in particular, that its  ground-state position (momentum) entropy increases (decreases) with the dimensionality in such a way that the sum $S_t^{(\mathtt{d})}$ exhibits practically linear dependence on $\mathtt{d}$ \cite{Yanez1}. After this, many other properties and asymptotic limits have been discussed too \cite{Angulo1,Yanez2,Dehesa4,LopezRosa1,Dehesa2,SanchezMoreno2,Rudnicki1,Toranzo4,Toranzo3,Toranzo7,PuertasCenteno1,Dehesa8,Toranzo1,Dehesa3,Toranzo2,Ikot1,Dehesa1,Aptekarev2}.

Contrary to the functionals $S_\rho^{(\mathtt{d})}$ and  $S_\gamma^{(\mathtt{d})}$, Fisher informations \cite{Fisher1,Frieden1}
\begin{subequations}\label{Fisher1}
\begin{align}\label{Fisher1_R}
I_{\rho_\mathtt{n}}^{(\mathtt{d})}&=\int_{\mathcal{D}_\rho^{(\mathtt{d})}}\rho_\mathtt{n}^{(\mathtt{d})}({\bf r})\left|{\bm\nabla}_{\bf r}\ln\rho_\mathtt{n}^{(\mathtt{d})}({\bf r})\right|^2\!\!d{\bf r}=\int_{\mathcal{D}_\rho^{(\mathtt{d})}}\frac{\left|{\bm\nabla}_{\bf r}\rho_\mathtt{n}^{(\mathtt{d})}({\bf r})\right|^2}{\rho_\mathtt{n}^{(\mathtt{d})}({\bf r})}d{\bf r}\\
\label{Fisher1_K}
I_{\gamma_\mathtt{n}}^{(\mathtt{d})}&=\int_{\mathcal{D}_\gamma^{(\mathtt{d})}}\gamma_\mathtt{n}^{(\mathtt{d})}({\bf k})\left|{\bm\nabla}_{\bf k}\ln\gamma_\mathtt{n}^{(\mathtt{d})}({\bf k})\right|^2\!\!d{\bf k}=\int_{\mathcal{D}_\gamma^{(\mathtt{d})}}\frac{\left|{\bm\nabla}_{\bf k}\gamma_\mathtt{n}^{(\mathtt{d})}({\bf k})\right|^2}{\gamma_\mathtt{n}^{(\mathtt{d})}({\bf k})}\,d{\bf k},
\end{align}
\end{subequations}
which contain gradients of the corresponding densities, are local measures of uncertainty with the extreme sensitivity to the speed of variation of the associated waveform. As it directly follows from Eqs.~\eqref{Fisher1}, its components,  independently of the dimensionality, are measured in units of the squared length (momentum part) or its inverse (position integral):
\begin{subequations}\label{FisherDimensionless1}
\begin{align}\label{FisherDimensionless1_rho}
I_{\rho_\mathtt{n}}^{(\mathtt{d})}&=L^{-2}\overline{I}_{\rho_\mathtt{n}}^{(\mathtt{d})},\\
\label{FisherDimensionless1_gamma}
I_{\gamma_\mathtt{n}}^{(\mathtt{d})}&=L^2\overline{I}_{\gamma_\mathtt{n}}^{(\mathtt{d})},
\end{align}
\end{subequations}
what makes their product a dimensionless quantity. Among the highly diverse applications of the Fisher information in science, engineering and technology \cite{Frieden1}, one has to stress its pivotal role in the density-functional theory where its position element defines the functional of the kinetic energy of the many-particle system  establishing in this way a relation between information and the kinetic energy \cite{Sears1} what allows, in particular, to reformulate the quantum mechanical variation principle as a precept of minimal information. The connections between the Fisher information and quantum mechanics also have wider implications for other branches of physics; for example, the constrained Fisher-optimization scheme leading to the Schr\"{o}dinger equation inspired the development of new approaches to aspects of nonequilibrium statistical mechanics \cite{Flego1}. Beyond physics, the functional $I$ is being actively used in, e.g., economics, analysis of cancer growth, transport processes, etc. \cite{Frieden1}. A remarkable example vividly exhibiting the avalanche-like growth of the interest to the Fisher measure and its applications in different fields is the fact that the first printing of the fundamental book, Ref.~\cite{Frieden1}, carried a title 'Physics from Fisher information' whereas in the second edition it has been changed to 'Science from Fisher information'. Let us also point out that originally this functional was introduced as a way of measuring the amount of information that an observable random variable $\cal X$ carries about an unknown parameter $\Theta$ upon which the probability $f({\cal X}|\Theta)$ depends, with $f({\cal X}|\Theta)$ being the probability density function for $\cal X$ conditioned on the value of $\Theta$. In this case, the Fisher information is defined by an expression involving the partial derivative of $\ln\!f$ with respect to the parameter $\Theta$ \cite{Fisher1}:
\begin{equation}\label{FisherOriginal1}
I_{\cal X}(\Theta)=\int_{\Omega_{{\cal X}|\Theta}}\!\!f({\cal X}|\Theta)\left[\frac{\partial\ln\!f({\cal X}|\Theta)}{\partial\Theta}\right]^2d{\cal X},
\end{equation}
with $\Omega_{{\cal X}|\Theta}$ being a set of all admissible values of $\cal X$. Instead, Eqs.~\eqref{Fisher1} involve the partial derivatives of the logarithm of the densities $\rho$ and $\gamma$ with respect to the spatial and wave vector coordinates, respectively, what means that our treatment is related to a particular case of Fisher information in which the parameter $\Theta$ corresponds to, say, the spatial shifts of the density $\rho$. Relevant to the main aim of the present research, let us mention that both position and momentum Fisher informations of the $\mathtt{d}$D hydrogenic-like systems are expressed in closed forms in terms of $\mathtt{d}$, the nuclear charge and the orbital quantum numbers \cite{Romera2,Dehesa6}. The same holds true for the harmonic oscillator with $\mathtt{d}\geq3$ where $I_{\rho_\mathtt{n}}^{(\mathtt{d})}$ and $I_{\gamma_\mathtt{n}}^{(\mathtt{d})}$ depend on the dimensionality of the space, confining strength and the indices of the corresponding state \cite{Dehesa1,Romera2}. Some other properties are addressed in Refs.~\cite{SanchezMoreno3,Dehesa5,Ikot1,Dehesa4,Romera1,SanchezMoreno1,Toranzo5,LopezRosa1,SobrinoColl1}.

Components $O_\rho^{(\mathtt{d})}$ and $O_\gamma^{(\mathtt{d})}$ of another quantum-mechanical measure that was proposed by O. Onicescu in 1966 \cite{Onicescu1} present averaged densities of the position or momentum densities and describe the deviation of the corresponding distribution from the uniform one:
\begin{subequations}\label{Onicescu1}
\begin{align}\label{Onicescu1_R}
O_{\rho_\mathtt{n}}^{(\mathtt{d})}&=\int_{\mathcal{D}_\rho^{(\mathtt{d})}}\left[\rho_\mathtt{n}^{(\mathtt{d})}({\bf r})\right]^2\!\!d{\bf r}\\
\label{Onicescu1_K}
O_{\gamma_\mathtt{n}}^{(\mathtt{d})}&=\int_{\mathcal{D}_\gamma^{(\mathtt{d})}}\left[\gamma_\mathtt{n}^{(\mathtt{d})}({\bf k})\right]^2\!\!d{\bf k}.
\end{align}
\end{subequations} 
These informational energies, as they were originally named by the Romanian mathematician in his publication introducing them \cite{Onicescu1} are inversely or directly proportional to the $\mathtt{d}$-th power of the characteristic length $L$:
\begin{subequations}\label{OnicescuDimensionless1}
\begin{align}\label{OnicescuDimensionless1_rho}
O_{\rho_\mathtt{n}}^{(\mathtt{d})}&=L^{-\mathtt{d}}\,\overline{O}_{\rho_\mathtt{n}}^{(\mathtt{d})},\\
\label{OnicescuDimensionless1_gamma}
O_{\gamma_\mathtt{n}}^{(\mathtt{d})}&=L^\mathtt{d}\,\overline{O}_{\gamma_\mathtt{n}}^{(\mathtt{d})}.
\end{align}
\end{subequations}
Accordingly, to prevent a possible confusion, it has to be underlined that they are not the energies in a regular meaning of this word, which are measured in Joules and are the eigenvalues of the Hamiltonian, but nevertheless, following the established tradition, they are commonly referred to as 'Onicescu energies'.

In an effort to generalize the Shannon entropy to the one-parameter functional that preserves its additivity, A. R\'{e}nyi introduced the measure that now bears his name \cite{Renyi1,Renyi2}. For the discrete set of all $N$ possible events with the probabilities $p_n$, $n=1,2,\ldots N$, with $0\leq p_n\leq1$ and $\sum_{n=1}^Np_n=1$, it is defined as
\begin{equation}\label{RenyiDiscrete1}
R_p(\alpha)=\frac{1}{1-\alpha}\ln\!\left(\sum_{n=1}^Np_n^\alpha\right),
\end{equation}
what for the continuous distributions $\rho^{(\mathtt{d})}({\bf r})$ and $\gamma^{(\mathtt{d})}({\bf k})$ transforms to
\begin{subequations}\label{Renyi1}
\begin{align}\label{Renyi1_R}
R_{\rho_\mathtt{n}}^{(\mathtt{d})}(\alpha)&=\frac{1}{1-\alpha}\ln\!\left(\int_{\mathcal{D}_\rho^{(\mathtt{d})}}\left[\rho_\mathtt{n}^{(\mathtt{d})}({\bf r})\right]^\alpha\!\!d{\bf r}\right)\\
\label{Renyi1_K}
R_{\gamma_\mathtt{n}}^{(\mathtt{d})}(\alpha)&=\frac{1}{1-\alpha}\ln\!\left(\!\int_{\mathcal{D}_\gamma^{(\mathtt{d})}}\left[\gamma_\mathtt{n}^{(\mathtt{d})}({\bf k})\right]^\alpha\!\!d{\bf k}\right).
\end{align}
\end{subequations}
R\'{e}nyi entropy is a decreasing function of its non-negative parameter, $0\leq\alpha<\infty$, which, in the limit of the unit $\alpha$, degenerates, according to the l'H\^{o}pital's rule, into its Shannon counterpart, $\lim_{\alpha\rightarrow1}R(\alpha)=S$.  From this point of view, $R(\alpha)$ can be construed as the measure of the sensitivity of the structure to its deviation from the equilibrium distribution, which is described by $\alpha=1$: for the very large factors, the events with the highest probability are the only contributors to the value of the entropy whereas the opposite regime of the extremely small $\alpha$ (provided it exists) treats all happenings on the same footing, independently of their actual occurrences. Also, as mentioned above, the R\'{e}nyi functional is additive, $R_{fg}(\alpha)=R_f(\alpha)+R_g(\alpha)$, and, similar to the Shannon entropies, for the continuous probabilities it is measured in the logarithm of length:
\begin{subequations}\label{RenyiDimensionless1}
\begin{align}\label{RenyiDimensionless1_rho}
R_{\rho_\mathtt{n}}^{(\mathtt{d})}(\alpha)&=\mathtt{d}\ln L+\overline{R}_{\rho_\mathtt{n}}^{(\mathtt{d})}(\alpha),\\
\label{RenyiDimensionless1_gamma}
R_{\gamma_\mathtt{n}}^{(\mathtt{d})}(\alpha)&=-\mathtt{d}\ln L+\overline{R}_{\gamma_\mathtt{n}}^{(\mathtt{d})}(\alpha).
\end{align}
\end{subequations}
Moreover, the R\'{e}nyi components of the two conjugate observables are not independent from each other but obey the following fundamental relation \cite{Bialynicki1,Zozor1}:
\begin{equation}\label{RenyiUncertainty1}
R_{\rho_\mathtt{n}}^{(\mathtt{d})}(\alpha)+R_{\gamma_\mathtt{n}}^{(\mathtt{d})}(\beta)\geq-\frac{\mathtt{d}}{2}\left(\frac{1}{1-\alpha}\ln\frac{\alpha}{\pi}+\frac{1}{1-\beta}\ln\frac{\beta}{\pi}\right),
\end{equation}
with the positive parameters $\alpha$ and $\beta$ being conjugated as
\begin{equation}\label{RenyiUncertainty2}
\frac{1}{\alpha}+\frac{1}{\beta}=2;
\end{equation}
in particular, at $\alpha\rightarrow1$, inequality~\eqref{RenyiUncertainty1} degenerates into its Shannon counterpart, Eq.~\eqref{ShannonInequality1}.

Non-additive one-parameter generalization of the Shannon measure is represented by the Tsallis \cite{Tsallis1} (or, more correctly from a historical point of view, Havrda-Charv\'{a}t-Dar\'{o}czy-Tsallis \cite{Havrda1,Daroczy1})  entropy that for the discrete events reads:
\begin{equation}\label{TsallisDiscrete1}
T_p(\alpha)=\frac{1}{\alpha-1}\left(1-\sum_{n=1}^Np_n^\alpha\right)
\end{equation}
with its continuous fellows being:
\begin{subequations}\label{Tsallis1}
\begin{align}\label{Tsallis1_R}
T_{\rho_\mathtt{n}}^{(\mathtt{d})}(\alpha)&=\frac{1}{\alpha-1}\left(1-\int_{\mathcal{D}_\rho^{(\mathtt{d})}}\left[\rho_\mathtt{n}^{(\mathtt{d})}({\bf r})\right]^\alpha\!\!d{\bf r}\right)\\
\label{Tsallis1_K}
T_{\gamma_\mathtt{n}}^{(\mathtt{d})}(\alpha)&=\frac{1}{\alpha-1}\left(1-\int_{\mathcal{D}_\gamma^{(\mathtt{d})}}\left[\gamma_\mathtt{n}^{(\mathtt{d})}({\bf k})\right]^\alpha\!\!d{\bf k}\right).
\end{align}
\end{subequations}
Its $\alpha=1$ limit brings the Tsallis entropy to the Shannon case, $\lim_{\alpha\rightarrow1}T(\alpha)=S$, but contrary to it and to the R\'{e}nyi configuration, it is only \textit{pseudo}-additive,
\begin{equation}\label{AdditivityTsallis1}
T_{fg}(\alpha)=T_f(\alpha)+T_g(\alpha)+(1-\alpha)T_f(\alpha)T_{g}(\alpha).
\end{equation}
Tsallis uncertainty relation \cite{Rajagopal1}
\begin{equation}\label{TsallisInequality1}
\left(\frac{\alpha}{\pi}\right)^{\mathtt{d}/(4\alpha)}\!\!\left[1+(1-\alpha)T_{\rho_\mathtt{n}}^{(\mathtt{d})}(\alpha)\right]^{1/(2\alpha)}\!\!\geq\!\!\left(\frac{\beta}{\pi}\right)^{\mathtt{d}/(4\beta)}\!\!\left[1+(1-\beta)T_{\gamma_\mathtt{n}}^{(\mathtt{d})}(\beta)\right]^{1/(2\beta)}
\end{equation}
is a direct consequence of the Sobolev inequality of the Fourier transform \cite{Beckner1}:
\begin{equation}\label{Sobolev1}
\left(\frac{\alpha}{\pi}\right)^{\mathtt{d}/(4\alpha)}\!\!\left(\int_{\mathcal{D}_\rho^{(\mathtt{d})}}\left[\rho_\mathtt{n}^{(\mathtt{d})}({\bf r})\right]^\alpha\!\!d{\bf r}\right)^{1/(2\alpha)}\!\!\geq\!\!\left(\frac{\beta}{\pi}\right)^{\mathtt{d}/(4\beta)}\!\!\left(\int_{\mathcal{D}_\gamma^{(\mathtt{d})}}\left[\gamma_\mathtt{n}^{(\mathtt{d})}({\bf k})\right]^\beta\!\!d{\bf k}\right)^{1/(2\beta)},
\end{equation}
and it holds true when, in addition to the conjugation from Equation~\eqref{RenyiUncertainty2}, an extra restriction
\begin{equation}\label{Sobolev2}
\frac{1}{2}\leq\alpha\leq1
\end{equation}
is imposed. At the right edge of the latter interval, the Tsallis, Eq.~\eqref{TsallisInequality1}, and Sobolev, Eq.~\eqref{Sobolev1}, relations turn into the identities. Note that the logarithmization of the Sobolev inequality from Eq.~\eqref{Sobolev1} results in the R\'{e}nyi uncertainty, Eq.~\eqref{RenyiUncertainty1}, simultaneously waiving for it the constraint from Eq.~\eqref{Sobolev2}. Uncertainty relations are an indispensable tool in data compression, quantum cryptography, entanglement witnessing, quantum metrology and other tasks employing correlations between the position and momentum components of the information measures \cite{Wehner1,Jizba2,Coles1,Toscano1,Hertz1,Wang1}. R\'{e}nyi and Tsallis entropies are expressed through each other as
\begin{subequations}\label{RenyiTsallisRelation1}
\begin{align}
\label{RenyiTsallisRelation1_1}
T&=\frac{1}{\alpha-1}\left[1-e^{(1-\alpha)R}\right]\\
\label{RenyiTsallisRelation1_2}
R&=\frac{1}{1-\alpha}\ln(1+(1-\alpha)T).
\end{align}
\end{subequations}
In addition, Onicescu energy, Eqs.~\eqref{Onicescu1}, is a particular case of them,
\begin{equation}\label{OnicescuRenyiTsallis1}
O=e^{-R(2)}=1-T(2).
\end{equation}
Second-order many-body R\'{e}nyi entanglement entropy of the Bose-Einstein condensates of the interacting atoms or ions was recently measured in state-of-the-art experiments \cite{Islam1,Kaufman1,Brydges1}. Properties of the R\'{e}nyi and Tsallis entropies are thoroughly addressed in many sources, see, e.g., Refs. \cite{Jizba1,Tsallis2,Tozzi1} and literature cited therein. Both of them are under very intensive scrutiny with immense applications in extremely diverse branches of the human activity, including their analysis for the $\mathtt{d}$D (with $\mathtt{d}\geq3$) quantum structures \cite{Dehesa7,LopezRosa1,Toranzo4,Dehesa8,Dehesa1,Aptekarev2,SobrinoColl1,Aptekarev1,Toranzo7,PuertasCenteno4,PuertasCenteno2,PuertasCenteno3,Ikot1}. As a final remark, one has to point out that the transition from the discrete events when the associated probabilities $p_n$ and, accordingly, Tsallis entropy, Eq.~\eqref{TsallisDiscrete1}, are dimensionless, to the continuous distribution leads to the dimensional incompatibility of the items in the right-hand sides of Eq.~\eqref{Tsallis1} what makes it impossible to study directly these two functionals but both parts of the uncertainty relation, Eq.~\eqref{TsallisInequality1}, are measured in the same units of $L^{\mathtt{d}(1-\alpha)/(2\alpha)}$, or, equivalently, $L^{\mathtt{d}(\beta-1)/(2\beta)}$, since, as it follows from Eq.~\eqref{RenyiUncertainty2},
\begin{equation}\label{Beta1}
\beta=\frac{\alpha}{2\alpha-1},
\end{equation}
and can be straightforwardly compared.

Recently, in an attempt to understand the influence of the boundary conditions (BCs) on the properties of all five mentioned above measures, an investigation has been carried out \cite{Olendski1}, which compared them for the circular 2D quantum dot of the radius $a$ with the Dirichlet edge requirement that zeroes the position function at the interface $\cal S$, $\left.\Psi^{(\mathtt{d})}({\bf r})\right|_{\cal S}=0$, and with the Neumann BC for which the normal derivative of the waveform $\Psi^{(\mathtt{d})}({\bf r})$ vanishes at the surface, $\left.\frac{\partial\Psi}{\partial {\bf n}}\right|_{\cal S}=0$, with $\bf n$ being a unit inward normal to $\cal S$. It was shown, in particular, that for any orbital the sum $S_t^{(\mathtt{2})}$ of the Shannon entropies, Eq.~\eqref{ShannonSum1}, is greater for the latter geometry what means that the switch from the Dirichlet to the Neumann BC decreases an overall knowledge about position and momentum of the quantum particle. Another crucial distinction is the different values of the lowest threshold of the R\'{e}nyi/Tsallis coefficient at which the momentum one-parameter functionals exist: for the Dirichlet configuration it is $2/5$ whereas for the Neumann BC it is equal to two thirds. Remarkably, for the 1D quantum well the Dirichlet critical value of one fourth is also smaller than its Neumann counterpart of one half \cite{Olendski2}. As the 2D Neumann threshold is greater than one half, its R\'{e}nyi uncertainty relation for the sum of the position and wave vector components $R_\rho(\alpha)+R_\gamma\left(\frac{\alpha}{2\alpha-1}\right)$ is valid in the range $[1/2,2)$ only with its logarithmic divergence at the right edge whereas for all other systems it is defined at any coefficient $\alpha$ not smaller than one half. It was stated (without proof) that for the $\mathtt{d}$D hyperspherical quantum structure these level-independent critical verges defining the bottom above which the functionals $R_\gamma^{(\mathtt{d})}(\alpha)$ and $T_\gamma^{(\mathtt{d})}(\alpha)$ exist, depend on the dimensionality as
\begin{subequations}\label{Ddimensional1}
\begin{align}\label{Ddimensional1_D}
\alpha_{TH}^D(\mathtt{d})&=\frac{\mathtt{d}}{\mathtt{d}+3}
\intertext{for the Dirichlet BC, and}
\label{Ddimensional1_N}\alpha_{TH}^N(\mathtt{d})&=\frac{\mathtt{d}}{\mathtt{d}+1}
\end{align}
\end{subequations}
- for the Neumann one. This, in turn, defines the range $[1/2,\alpha_R(\mathtt{d}))$ outside which the R\'{e}nyi uncertainty relation does not make sense as
\begin{subequations}\label{Ddimensional2}
\begin{align}\label{Ddimensional2_D}
\alpha_R^D(\mathtt{d})&=\frac{\mathtt{d}}{\mathtt{d}-3}\\
\label{Ddimensional2_N}\alpha_R^N(\mathtt{d})&=\frac{\mathtt{d}}{\mathtt{d}-1}.
\end{align}
\end{subequations}
As the momentum functional logarithmically diverges at the R\'{e}nyi parameter approaching from the right the limit $\alpha_{TH}$, the same is true for the sum $R_\rho^{(\mathtt{d})}(\alpha)+R_\gamma^{(\mathtt{d})}(\beta)$ at the coefficient $\alpha$ tending to $\alpha_R$ from the left. At the dimensionality unrestrictedly increasing, the limits from Eqs.~\eqref{Ddimensional1} and \eqref{Ddimensional2} come closer and closer to the Shannon case:
\begin{subequations}\label{Ddimenional3}
\begin{align}\label{Ddimenional3_1}
\alpha_{TH,R}^D(\mathtt{d}\rightarrow\infty)&\rightarrow1\mp\frac{3}{\mathtt{d}}+\frac{9}{\mathtt{d}^2}\mp\ldots\\
\label{Ddimenional3_2}
\alpha_{TH,R}^N(\mathtt{d}\rightarrow\infty)&\rightarrow1\mp\frac{1}{\mathtt{d}}+\frac{1}{\mathtt{d}^2}\mp\ldots.
\end{align}
\end{subequations}

Below, a rigorous comparative analysis of the $\mathtt{d}$D hypersperical quantum dot of the radius $a$ with the Dirichlet and Neumann BCs is provided. Physically, these two types of the restriction on the position function at the surface $\cal S$ are used for the description of  different materials and/or processes; for example, in solid-state physics the former one is relevant for the normal metals and semiconductors with the latter requirement pertinent to superconductors. The same applies, respectively, to the TM and TE modes inside electromagnetic waveguides and cavities with the Neumann BC being a standard requirement in acoustics, etc. A proof of the correctness of Eqs.~\eqref{Ddimensional1} is based on the derivation of the exact analytic form of the wave vector functions $\Phi^{(\mathtt{d})}({\bf k})$, which are directly obtained as the Fourier transforms of their position counterparts $\Psi^{(\mathtt{d})}({\bf r})$. Upon their substitution into the corresponding functionals, a comparison convergence test is applied to the integrals in  $R_\gamma^{(\mathtt{d})}(\alpha)$ and $T_\gamma^{(\mathtt{d})}(\alpha)$ leading in this way to Eqs.~\eqref{Ddimensional1}.  For either BC and arbitrary $\mathtt{d}$, the present configuration confirms the earlier statement \cite{Olendski1,Olendski2,Olendski3,Olendski4} that both uncertainty relations, Eqs.~\eqref{RenyiUncertainty1} and \eqref{TsallisInequality1}, for the ground orbital turn into the identities at $\alpha=1/2$. It is shown that at any dimensionality the sum of the Shannon entropies $S_t^{(\mathtt{d})}$ is smaller for the Dirichlet dot as compared to the Neumann BC. Contrary to the hydrogen atom \cite{Yanez1}, the ground-state momentum Shannon entropy $\overline{S}_\gamma^{(\mathtt{d})}$ increases with the dimensionality whereas its position fellow $\overline{S}_\rho^{(\mathtt{d})}$ has its maximum at $\mathtt{d}=3$ (Dirichlet BC) or $\mathtt{d}=5$ (Neumann case) after which it monotonically decreases. These features are explained by the description of the shape of the corresponding radial waveforms. In this way, the present research enriches the rapidly developing field of the analysis of the $\mathtt{d}$D systems which are of interest not only in atomic physics and chemistry, as exemplified by the references cited above (see also reviews \cite{Yaffe2,Chatterjee1}), but also, for example, in the quantum chromodynamics \cite{Witten1,Yaffe1} and in the theory of gravitation where the large $\mathtt{d}$ limit of Einstein's equations is used for describing the general aspects of the black holes and also allows extensions to problems in hydrodynamics, condensed matter physics, and nuclear physics \cite{Emparan1}. Reviews cited in the previous sentence contain a huge number of references from where a lot more information about the applications of the $\mathtt{d}$D models and their $1/\mathtt{d}$ expansion at $\mathtt{d}\rightarrow\infty$ can be retrieved. Regarding hyperspherical configuration, one has to especially mention that the large $\mathtt{d}$ limit of the "spherical model" \cite{Berlin1,Stanley1} has won an acceptance as a very good approximation of the Ising problem and realistic Heisenberg model \cite{Chatterjee1}.

Sect.~\ref{sec_WF1} derives analytic expressions of the position and momentum waveforms of the Dirichlet and Neumann hyperspherical dots. Based on it, Sect.~\ref{sec_Shannon1} discusses for either BC Shannon entropy, Fisher information and Onicescu energy whereas Sect.~\ref{sec_Renyi} focuses on the most characteristic features of the one-parameter functionals. Conclusions are summarized in Sect.~\ref{sec_Conclusions}.

\section{Wave functions}\label{sec_WF1}
A subject of our consideration is a $\mathtt{d}$D hyperspherical quantum dot of the radius $a$ what means that the spatial region $\mathcal{D}_\rho^{(\mathtt{d})}$ where the charged particle is free to move, $V({\bf r})\equiv0$, is defined by the requirement $r\leq a$ with the magnitude $r$ of the radius-vector $\bf r$ being$$r=\sqrt{\sum_{j=1}^\mathtt{d}x_j^2}.$$Due to the symmetry, a natural basis in which the analysis should be carried out is the system of the hyperspherical polar coordinates ${\bf r}=(r,\Omega_{{\bf r}_{\mathtt{d}-1}})$ \cite{Avery1}, where the angular part $\Omega_{{\bf r}_{\mathtt{d}-1}}$ comprises $\mathtt{d}-2$ spherical directions $\theta_{{\bf r}_1},\theta_{{\bf r}_2},\ldots,\theta_{{\bf r}_{\mathtt{d}-2}}$, and one polar dependence: $\Omega_{{\bf r}_{\mathtt{d}-1}}=(\theta_{{\bf r}_1},\theta_{{\bf r}_2},\ldots,\theta_{{\bf r}_{\mathtt{d}-2}},\varphi_{\bf r})$ with their ranges being $0\leq\theta_{{\bf r}_j}\leq\pi$, $j=1,2,\ldots,\mathtt{d}-2$, $0\leq\varphi_{\bf r}<2\pi$, whereas for the hyperball the radial component $r$ changes, as stated above, from zero to $a$. The relations between them and the Cartesian coordinates $x_1,x_2,\ldots,x_\mathtt{d}$ are:
\begin{align}
x_1&=r\sin\theta_{{\bf r}_1}\sin\theta_{{\bf r}_2}\cdots\sin\theta_{{\bf r}_{\mathtt{d}-2}}\cos\varphi_{\bf r}\nonumber\\
x_2&=r\sin\theta_{{\bf r}_1}\sin\theta_{{\bf r}_2}\cdots\sin\theta_{{\bf r}_{\mathtt{d}-2}}\sin\varphi_{\bf r}\nonumber\\
x_3&=r\sin\theta_{{\bf r}_1}\sin\theta_{{\bf r}_2}\cdots\cos\theta_{{\bf r}_{\mathtt{d}-2}}\nonumber\\
\vdots&\nonumber\\
x_{\mathtt{d}-1}&=r\sin\theta_{{\bf r}_1}\cos\theta_{{\bf r}_2}\nonumber\\
x_\mathtt{d}&=r\cos\theta_{{\bf r}_1}.\nonumber
\end{align}
The Schr\"{o}dinger equation inside the dot transforms to
\begin{equation}\label{Schrodinger2}
-\frac{\hbar^2}{2m^*}\left(\frac{\partial^2}{\partial r^2}+\frac{\mathtt{d}-1}{r}\frac{\partial}{\partial r}-\frac{\widehat{\bf\Lambda}^2}{r^2}\right) \Psi_{n,l,\left\{\mu\right\}}^{(\mathtt{d})}({\bf r})=E_{n,l}^{(\mathtt{d})}\Psi_{n,l,\left\{\mu\right\}}^{(\mathtt{d})}({\bf r}),
\end{equation}
where the hyperindex $\mathtt{n}$ has been expanded for the present geometry as a set of the radial quantum number $n$, which is a positive integer, $n=1,2,\ldots$, and $\mathtt{d}-1$ angular indices $l,\left\{\mu\right\}$ with the orbital quantum number $l\equiv\mu_1$ and magnetic quantum numbers $\left\{\mu\right\}\equiv\mu_2,\mu_3,\ldots,\mu_{\mathtt{d}-1}$, such that $$l\geq\mu_2\geq\mu_3\geq\cdots\geq\left|\mu_{\mathtt{d}-1}\right|,$$
$\mu_{\mathtt{d}-1}\equiv m$. The integers $l,\mu_2,\cdots,\mu_{\mathtt{d}-2}$ are natural numbers, $l,\mu_2,\cdots,\mu_{\mathtt{d}-2}\in\mathbb{N}$, and $m$ can take negative values, $m\in\mathbb{Z}$. The radius-independent operator $\hbar^2\widehat{\bf\Lambda}^2$ in Eq.~\eqref{Schrodinger2} is a square of generalized angular momentum \cite{Avery1,Gallup1,Louck1,Chatterjee1}:
\begin{equation}\label{AngularMomentum1}
\widehat{\bf\Lambda}^2=-\!\!\sum_{\tiny\begin{array}{c}
i,j=1\\
i>j
\end{array}}^\mathtt{d}\!\!\left(x_i\frac{\partial}{\partial x_j}-x_j\frac{\partial}{\partial x_i}\right)^2\!\!=\!-\sum_{i=1}^{\mathtt{d}-1}\!\frac{\sin^{i+1-\mathtt{d}}\theta_i}{\Pi_{j=1}^{i-1}\sin^2\theta_j}\frac{\partial}{\partial\theta_i}\!\!\left(\!\sin^{\mathtt{d}-i-1}\theta_i\frac{\partial}{\partial\theta_i}\!\right),
\end{equation}
which in the hyperspherical coordinates satisfies the eigenvalue equation \cite{Avery1,Gallup1,Louck1,Chatterjee1}:
\begin{equation}\label{AngularEigenEquation1}
\widehat{\bf\Lambda}^2{\cal Y}_{l,\left\{\mu\right\}}^{(\mathtt{d})}(\Omega_{\mathtt{d}-1})=l(l+\mathtt{d}-2){\cal Y}_{l,\left\{\mu\right\}}^{(\mathtt{d})}(\Omega_{\mathtt{d}-1}).
\end{equation}
Here, the orthonormalized,
\begin{equation}\label{AngularOrthoNormal1}
\int{{\cal Y}_{l,\left\{\mu\right\}}^{(\mathtt{d})}}^{\!\!\!\!\!*}(\Omega_{\mathtt{d}-1}){\cal Y}_{l',\left\{\mu'\right\}}^{(\mathtt{d})}(\Omega_{\mathtt{d}-1})d\Omega_{\mathtt{d}-1}=\delta_{ll'}\delta_{\left\{\mu\right\},\left\{\mu'\right\}},
\end{equation}
hyperspherical harmonics ${\cal Y}_{l,\left\{\mu\right\}}^{(\mathtt{d})}(\Omega_{\mathtt{d}-1})$ are defined as \cite{Avery1}
\begin{equation}\label{AngularFunction1}
{\cal Y}_{l,\left\{\mu\right\}}^{(\mathtt{d})}(\Omega_{\mathtt{d}-1})={\cal N}_{l,\left\{\mu\right\}}e^{im\varphi}\Pi_{j=1}^{\mathtt{d}-2}C_{\mu_j-\mu_{j+1}}^{\gamma_j+\mu_{j+1}}(\cos\theta_j)\sin^{\mu_{j+1}}\!\theta_j,
\end{equation}
with $\gamma_j=(\mathtt{d}-j-1)/2$, $C_n^\lambda(z)$ is the Gegenbauer polynomial \cite{Olver1}, and normalization constant ${\cal N}_{l,\left\{\mu\right\}}$ reads:
$${\cal N}_{l,\left\{\mu\right\}}=\sqrt{\frac{1}{2\pi}\Pi_{j=1}^{\mathtt{d}-2}\frac{(\gamma_j+\mu_j)\Gamma(\gamma_j+\mu_{j+1})(\mu_j-\mu_{j+1})!(2\gamma_j+2\mu_{j+1}-1)!}{\pi^{1/2}\Gamma\left(\gamma_j+\mu_{j+1}+\frac{1}{2}\right)(2\gamma_j+\mu_j+\mu_{j+1}-1)!}},$$ $\Gamma(x)$ is $\Gamma$-function \cite{Olver1}. Separation of variables
\begin{equation}\label{Separation1}
\Psi_{n,l,\left\{\mu\right\}}^{(\mathtt{d})}({\bf r})={\cal R}_{n,l}^{(\mathtt{d})}(r){\cal Y}_{l,\left\{\mu\right\}}^{(\mathtt{d})}\left(\Omega_{{\bf r}_{\mathtt{d}-1}}\right)
\end{equation}
leads to the following equation for the radial dependence ${\cal R}_{n,l}^{(\mathtt{d})}(r)$:
\begin{equation}
\left[\frac{d^2}{dr^2}+\frac{\mathtt{d}-1}{r}\frac{d}{dr}+\frac{2m^*E_{n,l}^{(\mathtt{d})}}{\hbar^2}-\frac{l(l+\mathtt{d}-2)}{r^2}\right]{\cal R}_{n,l}^{(\mathtt{d})}(r)=0,
\end{equation}
whose orthonorlamized,
\begin{equation}\label{RadialOrthonormal1}
\int_0^a{\cal R}_{n',l}^{(\mathtt{d})}(r){\cal R}_{n,l}^{(\mathtt{d})}(r)r^{\mathtt{d}-1}dr=\delta_{nn'},
\end{equation}
solutions strongly depend on the function behavior at the boundary. If it vanishes at the hypersurface of the dot, ${\cal R}_{n,l}^{(\mathtt{d})D}(a)=0$ (Dirichlet BC), then
\begin{subequations}\label{DirichletPositionEnergy1}
\begin{align}\label{DirichletPosition1}
{\cal R}_{n,l}^{(\mathtt{d})D}(r)&=\frac{2^{1/2}}{a^{\mathtt{d}/2}}\frac{1}{j_{l+1}^{(\mathtt{d})}(j_{l+\mathtt{d}_1,n})}j_l^{(\mathtt{d})}\left(j_{l+\mathtt{d}_1,n}\frac{r}{a}\right)\\
\label{DirichletEnergy1}
E_{n,l}^{(\mathtt{d})D}&=\frac{\hbar^2}{2m^*a^2}j_{l+\mathtt{d}_1,n}^2,
\end{align}
\end{subequations}
and for the Neumann edge requirement, $\left.\frac{d}{dr}{\cal R}_{n,l}^{(\mathtt{d})N}(r)\right|_{r=a}=0$, one has:
\begin{subequations}\label{NeumannPositionEnergy1}
\begin{align}\label{NeumannPosition1}
{\cal R}_{n,l}^{(\mathtt{d})N}(r)&=\left\{\begin{array}{cc}
\frac{\mathtt{d}^{1/2}}{a^{\mathtt{d}/2}},&n=1,\,l=0\\
\frac{2^{1/2}}{a^{\mathtt{d}/2}}\frac{a_{l,n}^{(\mathtt{d})}}{\left(\left[a_{l,n}^{(\mathtt{d})}\right]^2-l(l+\mathtt{d}-2)\right)^{1/2}}\frac{1}{j_l^{(\mathtt{d})}\left(a_{l,n}^{(\mathtt{d})}\right)}j_l^{(\mathtt{d})}\!\left(a_{l,n}^{(\mathtt{d})}\frac{r}{a}\right),&{\rm all\,\,other\,\,cases}
\end{array}\right.\\
\label{NeumannEnergy1}
E_{n,l}^{(\mathtt{d})N}&=\frac{\hbar^2}{2m^*a^2}\left[a_{l,n}^{(\mathtt{d})}\right]^2.
\end{align}
\end{subequations}
Here, $j_l^{(\mathtt{d})}(z)$ is a hypersphrerical Bessel function \cite{Avery1}:
\begin{equation}\label{HyperBessel1}
j_l^{(\mathtt{d})}(z)=\frac{2^{\mathtt{d}_1-1}\Gamma(\mathtt{d}_1)}{(\mathtt{d}-4)!!z^{\mathtt{d}_1}}J_{\mathtt{d}_1+l}(z),
\end{equation}
$\mathtt{d}_1=\mathtt{d}/2-1$, $J_\nu(z)$ is a $\nu$th order Bessel function of the first kind \cite{Olver1}, $j_{\nu,n}$ is its $n$th zero, $J_\nu(j_{\nu,n})=0$, and $a_{l,n}^{(\mathtt{d})}$ is $n$th zero of the derivative of the $l$th order hyperspherical Bessel function, ${j_l^{(\mathtt{d})}}'\!\left(a_{l,n}^{(\mathtt{d})}\right)=0$. Note that at any dimensionality, $a_{0,1}^{(\mathtt{d})}\equiv0$.

Knowledge of the position waveforms paves the way to the calculation of their wave vector counterparts in the corresponding {\bf k} space $\mathcal{D}_\gamma^{(\mathtt{d})}=(k,\Omega_{{\bf k}_{\mathtt{d}-1}})\equiv(k,\theta_{{\bf k}_1},\theta_{{\bf k}_2},\ldots,\theta_{{\bf k}_{\mathtt{d}-2}},\varphi_{\bf k})$, $0\leq k<\infty$. Since the plane-wave expansion in the $\mathtt{d}$ dimensions is represented as \cite{Avery1}
\begin{equation}\label{PlaneWave1}
e^{i{\bf k}{\bf r}}=(\mathtt{d}-2)!!\frac{2\pi^{\mathtt{d}/2}}{\Gamma(\mathtt{d}/2)}\sum_{l=0}^\infty i^lj_l^{(\mathtt{d})}(kr)\sum_{\left\{\mu\right\}}{{\cal Y}_{l,\left\{\mu\right\}}^{(\mathtt{d})}}^{\!\!\!\!\!*}(\Omega_{{\bf k}_{\mathtt{d}-1}}){\cal Y}_{l,\left\{\mu\right\}}^{(\mathtt{d})}(\Omega_{{\bf r}_{\mathtt{d}-1}}),
\end{equation}
the Fourier transform, Eq.~\eqref{Fourier1_1}, preserves a separation of variables in such a way that the shape of the angular dependence stays intact:
\begin{equation}\label{Separation2}
\Phi_{n,l,\left\{\mu\right\}}^{(\mathtt{d})}({\bf k})={\cal K}_{n,l}^{(\mathtt{d})}(k){\cal Y}_{l,\left\{\mu\right\}}^{(\mathtt{d})}\left(\Omega_{{\bf k}_{\mathtt{d}-1}}\right),
\end{equation}
where the radial function reads:
\begin{equation}\label{RadialK1}
{\cal K}_{n,l}^{(\mathtt{d})}(k)=(-i)^l\frac{(\mathtt{d}-2)!!}{2^{\mathtt{d}_1}\Gamma(\mathtt{d}/2)}\int_0^a{\cal R}_{n,l}^{(\mathtt{d})}(r)j_l^{(\mathtt{d})}(kr)r^{\mathtt{d}-1}dr.
\end{equation}
Obviously, the position orthonormality, Eq.~\eqref{RadialOrthonormal1}, is inherited by the wave vector space too:
\begin{equation}\label{RadialOrthonormal2}
\int_0^\infty{\cal K}_{n',l}^{(\mathtt{d})}(k){\cal K}_{n,l}^{(\mathtt{d})}(k)k^{\mathtt{d}-1}dk=\delta_{nn'}.
\end{equation}
In deriving these equations, the $\mathtt{d}$D volume elements $d{\bf r}=r^{\mathtt{d}-1}drd\Omega_{{\bf r}_{\mathtt{d}-1}}$ and $d{\bf k}=k^{\mathtt{d}-1}dkd\Omega_{{\bf k}_{\mathtt{d}-1}}$ have been used, where $d\Omega_{\mathtt{d}-1}=d\varphi\Pi_{j=1}^{\mathtt{d}-2}\sin^{\mathtt{d}-j-1}\theta_jd\theta_j$. Explicit evaluation \cite{Prudnikov2} yields:
\begin{subequations}\label{MomentumFunction1}
\begin{align}\label{MomentumFunction1D}
{\cal K}_{n,l}^{(\mathtt{d})D}(k)&=(-i)^la^{\mathtt{d}/2}2^{(3-\mathtt{d})/2}\frac{(\mathtt{d}-2)!!}{\Gamma(\mathtt{d}/2)}\frac{j_{l+\mathtt{d}_1,n}}{j_{l+\mathtt{d}_1,n}^2-(ak)^2}j_l^{(\mathtt{d})}(ak)\\
\label{MomentumFunction1N}
{\cal K}_{n,l}^{(\mathtt{d})N}(k)&=\left\{\begin{array}{cc}
a^{\mathtt{d}/2}\mathtt{d}^{1/2}\frac{(\mathtt{d}-2)!!}{2^{\mathtt{d}_1}\Gamma(\mathtt{d}/2)}\frac{1}{ak}j_1^{(\mathtt{d})}(ak),&n=1,\,l=0\\
(-i)^la^{\mathtt{d}/2}2^{(3-\mathtt{d})/2}\frac{(\mathtt{d}-2)!!}{\Gamma(\mathtt{d}/2)}\frac{a_{l,n}^{(\mathtt{d})}}{\left(\left[a_{l,n}^{(\mathtt{d})}\right]^2-l(l+\mathtt{d}-2)\right)^{1/2}}\frac{ak}{\left[a_{l,n}^{(\mathtt{d})}\right]^2-(ak)^2}{j_l^{(\mathtt{d})}}'\!\!(ak),&{\rm all\,\,other\,\,cases,}
\end{array}\right.
\end{align}
\end{subequations}
where the derivative of the hyperspherical Bessel function ${j_l^{(\mathtt{d})}}'(z)$ satisfies the same recurrence relation
\begin{equation}
{j_l^{(\mathtt{d})}}'(z)=-j_{l+1}^{(\mathtt{d})}(z)+\frac{l}{z}j_l^{(\mathtt{d})}(z),
\end{equation}
as its regular counterpart $J_\nu(z)$ \cite{Olver1}. For the 3D spherical structure, the waveforms are \cite{Olendski1}:

for the Dirichlet BC:
\begin{subequations}\label{Function3D}
\begin{align}\label{Function3D_PositionDirichlet}
\Psi_{nlm}^{(\mathtt{3})D}(r,\theta_{\bf r},\varphi_{\bf r})&=\frac{2^{1/2}}{a^{3/2}j_{l+1}(j_{l+1/2,n})}j_l\!\left(j_{l+1/2,n}\frac{r}{a}\right)Y_{lm}(\theta_{\bf r},\varphi_{\bf r})\\
\label{Function3D_MomentumDirichlet}
\Phi_{nlm}^{(\mathtt{3})D}(k,\theta_{\bf k},\varphi_{\bf k})&=(-i)^la^{3/2}\frac{2}{\pi^{1/2}}\frac{j_{l+1/2,n}}{j_{l+1/2,n}^2-(ak)^2} j_l(ak)Y_{lm}(\theta_{\bf k},\varphi_{\bf k});\\
\intertext{for the Neumann requirement:}
\label{Function3D_PositionNeumann}
\Psi_{nlm}^{(\mathtt{3})N}(r,\theta_{\bf r},\varphi_{\bf r})&=\left\{
\begin{array}{cc}
\frac{1}{a^{3/2}}\left(\frac{3}{2\pi}\right)^{1/2},&n=1,\,\,l=0\\
\frac{2^{1/2}}{a^{3/2}}\frac{a_{l,n}'}{\left[a_{l,n}'^{\,2}-l(l+1)\right]^{1/2}}\frac{j_l\!\left(a_{l,n}'\frac{r}{a}\right)}{j_l(a_{l,n}')}Y_{lm}(\theta_{\bf r},\varphi_{\bf r}),&{\rm all\,\,other\,\,cases}
\end{array}\right.\\
\label{Function3D_MomentumNeumann}
\Phi_{nlm}^{(\mathtt{3})N}(k,\theta_{\bf k},\varphi_{\bf k})&=
\left\{\begin{array}{cc}
a^{3/2}\frac{3^{1/2}}{\pi}\frac{1}{ak}{j_1}(ak),&n=1,\,\,l=0\\
(-i)^la^{3/2}\frac{2}{\pi^{1/2}}\frac{a_{l,n}'}{\left[a_{l,n}'^{\,2}-l(l+1)\right]^{1/2}}\frac{ak}{a_{l,n}'^{\,2}-(ak)^2}{j_l}'(ak)Y_{lm}(\theta_{\bf k},\varphi_{\bf k}),&{\rm all\,\,other\,\,cases}.
\end{array}
\right.
\end{align}
\end{subequations}
Here, $Y_{lm}(\theta,\varphi)$ are standard  orthonormalized spherical harmonics \cite{Landau1}, $j_l(z)=\sqrt{\pi/(2z)}J_{l+1/2}(z)$ is the spherical Bessel function \cite{Olver1}, and $a_{l,n}'$ is the value at which its derivative turns to zero, $j_l'(a_{l,n}')=0$ \cite{Olver1}.

\section{Shannon entropy, Fisher information and Onicescu energy}\label{sec_Shannon1}
For the central potentials, due to the separation of variables, Eqs.~\eqref{Separation1} and \eqref{Separation2}, the logarithm in the Shannon functionals, Eqs.~\eqref{Shannon1}, splits either of them into the sum of the radial $S_{rad}$ and angular $S_{ang}$ entropies
\begin{equation}\label{ShannonSum2}
S=S_{rad}+S_{ang},
\end{equation}
with the latter being the same for the position and momentum components. Properties of $S_{ang}$ were studied before \cite{Yanez2,Dehesa4}. Similarly, expressions for the angular parts of the Fisher information and Onicescu energy are also known \cite{Dehesa4}. Accordingly, here we stay focused on the ground state, $n=1$, $l=0$, when the angular density reduces to just the $\theta$- and $\varphi$-independent constant:
\begin{equation}\label{Angular1}
{{\cal Y}_{0,\left\{0\right\}}^{(\mathtt{d})}}^{\!\!\!\!\!2}\left(\Omega_{\mathtt{d}-1}\right)=\frac{1}{2\pi}\Pi_{j=1}^{\mathtt{d}-2}\frac{\Gamma\left(\frac{\mathtt{d}-j+1}{2}\right)}{\pi^{1/2}\Gamma\left(\frac{\mathtt{d}-j}{2}\right)}.
\end{equation}
Our primary subject of interest will be the entropy (and other measures) dependence on the dimensionality and BC. We will also draw parallels to the other central potential dependencies $V({\bf r})$.

\begin{figure}
\centering
\includegraphics[width=\columnwidth]{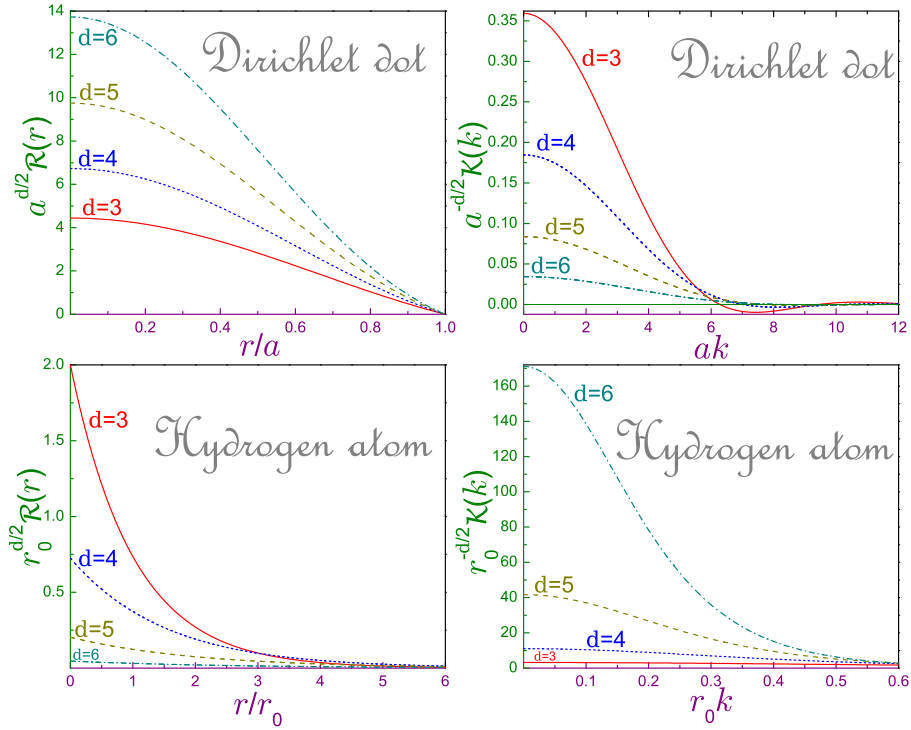}
\caption{\label{Fig_Waveforms}
Upper panels show dimensionless Dirichlet radial ground-state position $a^{\mathtt{d}/2}{\cal R}_{1,0}^{(\mathtt{d})D}(r)$ (left window) and momentum $a^{-\mathtt{d}/2}{\cal K}_{1,0}^{(\mathtt{d})D}(k)$ (right subplot) functions in terms of the unitless distance $r/a$ and wave vector $ak$, respectively. The lower pictures exhibit the same dependencies for the hydrogen atom with the dimensionless length and wave vector being $r/r_0$ and $r_0k$, respectively. Solid lines are for the 3D geometry, dotted ones  - for the dimensionality $\mathtt{d}=4$, dashed curves are for $\mathtt{d}=5$ and dash-dotted characteristics are for the 6D structures.}
\end{figure}

Table~\ref{Table_Dirichlet} shows dimensionless position and wave vector Shannon, Fisher and Onicescu measures together with their sum (Shannon) or product (Fisher and Onicescu) for the Dirichlet BC with Table~\ref{Table_Neumann} exhibiting the results for the Neumann dot. Obviously, the natural unit of length for the present geometry is the dot hyperradius, $L\equiv a$. It is seen that the momentum Shannon entropy monotonically increases with the dimensionality whereas its position counterpart possesses a maximum at $\mathtt{d}=3$ ($\mathtt{d}=5$) for the Dirichlet (Neumann) BC. Such behavior is opposite to that of the hydrogen atom \cite{Yanez1}. To explain the difference, one has to consider the corresponding waveforms. For the hydrogen-like ion with the potential 
\begin{equation}\label{HydrogenPotential1}
V({\bf r})=-Zk_e\frac{e^2}{r},
\end{equation}
where $Z$ is the number of protons in the nucleus, $e$ is an absolute value of the electronic charge, $k_e=1/(4\pi\varepsilon_0)$ is the Coulomb constant, $\varepsilon_0=8.854\ldots\times10^{-12}$ F/m is the vacuum permittivity, the radial dependencies are \cite{Yanez1,Nieto1,Aquilanti1}:
\begin{subequations}\label{Hydrogen1}
\begin{align}
{\cal R}_{n,l}^{(\mathtt{d})H}(r)&=\frac{1}{(\lambda r_0)^{\mathtt{d}/2}}\left[\frac{(n-l-1)!}{4\lambda(n+l+\mathtt{d}-3)!}\right]^{1/2}\nonumber\\
\label{Hydrogen1_Pos}
\times&\exp\!\left(\!-\frac{r}{2\lambda r_0}\right)\!\!\left(\frac{r}{\lambda r_0}\right)^l\!\!L_{n-l-1}^{(2l+\mathtt{d}-2)}\!\!\left(\!\frac{r}{\lambda r_0}\!\right),\quad 0\leq r<\infty\\
{\cal K}_{n,l}^{(\mathtt{d})H}(k)&=(2\lambda r_0)^{\mathtt{d}/2}2^{2l+\mathtt{d}}\left[\frac{\lambda(n-l-1)!}{\pi(n+l+\mathtt{d}-3)!}\right]^{1/2}\Gamma\!\left(l+\frac{\mathtt{d}-1}{2}\right)\nonumber\\
\label{Hydrogen1_Mom}
\times&\frac{(2\lambda r_0k)^l}{\left[1+(2\lambda r_0k)^2\right]^{l+(\mathtt{d}+1)/2}}C_{n-l-1}^{(l+(\mathtt{d}-1)/2)}\!\!\left(\frac{1-(2\lambda r_0k)^2}{1+(2\lambda r_0k)^2}\right).
\end{align}
\end{subequations}
Here, $r_0=a_0/Z$ is a characteristic length of the ion that is expressed via the Bohr radius $a_0=\hbar^2/(m_ek_ee^2)$, $m_e$ is electronic mass, $\lambda=\frac{1}{2}\left(n+\frac{\mathtt{d}-3}{2}\right)$, $L_n^{(\eta)}(z)$ is generalized Laguerre polynomial \cite{Olver1}, and orbital quantum number is bounded from above, $l=0,1,\ldots,n-1$. As Fig.~\ref{Fig_Waveforms} shows, the increase of the dimensionality leads to the stronger concentration of the electron around the centre of the Dirichlet dot whereas the momentum distribution is subdued and gets more uniform what means that our knowledge (ignorance) about the particle position (wave vector) enlarges. Then, according to the interpretation of the Shannon entropy provided in the Introduction, $S_\rho^{(\mathtt{d})D}$ decreases and $S_\gamma^{(\mathtt{d})D}$ increases with the dimensionality and, as Tables~\ref{Table_Dirichlet} and \ref{Table_Neumann} exhibit, these variations cause the sum $S_t^{(\mathtt{d})}$ of the two to grow practically linearly with $\mathtt{d}$. Consulting Fig.~\ref{Fig_Waveforms} again, for the hydrogen atom one observes the opposite tendency: the growing dimensionality smooths out  the position distribution making it more and more homogeneous what means a decrease of our knowledge and the rise of the related entropy $S_\rho^{(\mathtt{d})H}$. Simultaneously, the probability of finding the corpuscle with the zero momentum  increases shrinking in this way the corresponding uncertainty and bringing more information about the motion. Note that the sum of the two hydrogen entropies, similar to the both BC types of the quantum dot, almost does not deviate from the linear dependence on $\mathtt{d}$ \cite{Yanez1}. Negative values of the Shannon entropy for either surface requirement, which in the case of the discrete events
\begin{equation}\label{ShannonDiscrete1}
S=-\sum_{n=1}^Np_n\ln p_n
\end{equation}
is always positive, is explained by the superiority of the contribution from the regions where the absolute value of the distribution function is greater than unity \cite{Rudnicki2}. Since, at the fixed $\mathtt{d}$, the Dirichlet position as well as momentum components are smaller than their Neumann counterparts, the same holds true for the sum, Eq.~\eqref{ShannonSum1}, that enters the corresponding uncertainty relation, Eq.~\eqref{ShannonInequality1}, which is always satisfied, as a comparison of the corresponding columns in Tables~\ref{Table_Dirichlet} and \ref{Table_Neumann} demonstrates. Thus, the Neumann BC provides less information about both the particle position as well as momentum and, accordingly, for it the total ignorance of the motion is greater: the Dirichlet value of $S_\rho^{(\mathtt{d})}+S_\gamma^{(\mathtt{d})}$ lies much closer to the fundamental limit $\mathtt{d}(1+\ln\pi)$ than the Neumann one. Let us also note here that due to the simplicity of the ground-state position Neumann waveform, Eq.~\eqref{NeumannPosition1}, the corresponding measures can be elementary calculated:
\begin{subequations}\label{NeumannPositionMeasures1}
\begin{align}\label{NeumannPositionMeasures1_Shannon}
S_{\rho_{1,0,\left\{0\right\}}}^{(\mathtt{d})N}&=\mathtt{d}\ln a-\ln\!\left(\frac{\mathtt{d}}{2\pi}\Pi_{j=1}^{\mathtt{d}-2}\frac{\Gamma\left(\frac{\mathtt{d}-j+1}{2}\right)}{\pi^{1/2}\Gamma\left(\frac{\mathtt{d}-j}{2}\right)}\right)\\
\label{NeumannPositionMeasures1_Fisher}
I_{\rho_{1,0,\left\{0\right\}}}^{(\mathtt{d})N}&=0\\
\label{NeumannPositionMeasures1_Onicescu}
O_{\rho_{1,0,\left\{0\right\}}}^{(\mathtt{d})N}&=\frac{1}{a^\mathtt{d}}\frac{\mathtt{d}}{2\pi}\Pi_{j=1}^{\mathtt{d}-2}\frac{\Gamma\left(\frac{\mathtt{d}-j+1}{2}\right)}{\pi^{1/2}\Gamma\left(\frac{\mathtt{d}-j}{2}\right)}.
\end{align}
\end{subequations}

For the central potentials, $V({\bf r})\equiv V(r)$, expressions for the Fisher informations can be brought to the forms \cite{Romera1}:
\begin{subequations}\label{FisherCentral1}
\begin{align}\label{FisherCentral1_R}
I_\rho^{(\mathtt{d})}&=4\left\langle k^2\right\rangle-2|m|(2l+\mathtt{d}-2)\left\langle r^{-2}\right\rangle\\
\label{FisherCentral1_K}
I_\gamma^{(\mathtt{d})}&=4\left\langle r^2\right\rangle-2|m|(2l+\mathtt{d}-2)\left\langle k^{-2}\right\rangle
\end{align}
\end{subequations}
with $\left\langle\ldots\right\rangle$ denoting quantum-mechanical averaging. Using further simplifications for $\left\langle k^2\right\rangle$ \cite{Romera1}:
\begin{equation}\label{K2averaging1}
\left\langle k^2\right\rangle=\int_0^a\left(\frac{d{\cal R}_{n,l}^{(\mathtt{d})}(r)}{dr}\right)^2r^{\mathtt{d}-1}dr+l(l+\mathtt{d}-2)\left\langle r^{-2}\right\rangle,
\end{equation}
one can correspondingly rewrite Eq.~\eqref{FisherCentral1_R}. In general, the integrals in the last equation, even though can be calculated analytically, are not very transparent since they contain (for the odd dimensionalities) the values of the derivative of the Bessel functions $J_\nu(z)$ with respect to index $\nu$ and infinite series of the Bessel functions or, at best, (for the even $\mathtt{d}$) finite sums of $J_\nu(z)$ \cite{Prudnikov2}. However, for some particular orbitals, position and momentum Fisher informations can be obtained in quite compact form; for example, for the spherically symmetric states, the position  functionals read:
\begin{subequations}\label{FisherCases1}
\begin{align}\label{FisherCases1a}
I_{\rho_{n,0,\left\{0\right\}}}^{(\mathtt{d})D}&=4\frac{j_{\mathtt{d}_1,n}^2}{a^2}\\
\label{FisherCases1b}
I_{\rho_{n,0,\left\{0\right\}}}^{(\mathtt{d})N}&=4\frac{\left[a_{0,n}^{(\mathtt{d})}\right]^2}{a^2},
\intertext{$n=1,2,\ldots$, what makes them proportional to the corresponding energies, Eqs.~\eqref{DirichletEnergy1} and \eqref{NeumannEnergy1}:}
\label{FisherCases1c}
I_{\rho_{n,0,\left\{0\right\}}}^{(\mathtt{d})}&=\frac{8m^*}{\hbar^2}E_{n,0}.
\intertext{Note that for the Dirichlet 1D well and 3D dot the root $j_{\mathtt{d}_1,n}$ is the same causing the identical informations, as the corresponding entries of Table~\ref{Table_Dirichlet} manifest. Also, some momentum measures are:}
\label{FisherCases1d}
I_{\gamma_{n,0,0}}^{(\mathtt{3})D}&=\frac{2}{3}\frac{2n^2\pi^2-3}{n^2\pi^2}a^2\\
\label{FisherCases1e}
I_{\gamma_{n,0,0,0}}^{(\mathtt{4})D}&=I_{\gamma_{n+1,0,0,0}}^{(\mathtt{4})N}=\frac{4}{3}a^2\\
\label{FisherCases1f}
I_{\gamma_{1,0,\left\{0\right\}}}^{(\mathtt{d})N}&=4\frac{\mathtt{d}}{\mathtt{d}+2}a^2.
\end{align}
\end{subequations}
Eq.~\eqref{FisherCases1d} means that the high-lying spherically symmetric 3D Dirichlet orbitals saturate momentum Fisher information to $\frac{4}{3}a^2$:
\begin{subequations}
\begin{align}\tag{58d$'$}\label{eq:58d'}
I_{\gamma_{n,0,0}}^{(\mathtt{3})D}&\rightarrow\left(\frac{4}{3}-\frac{2}{n^2\pi^2}\right)a^2,\quad n\rightarrow\infty,
\intertext{whereas the unlimited increase of the dimensionality enforces the Neumann ground-state momentum Fisher functional to approach the value of $4a^2$:}
\tag{58f$'$}\label{eq:58f'}
I_{\gamma_{1,0,\left\{0\right\}}}^{(\mathtt{d})N}&\rightarrow4\left(1-\frac{2}{\mathtt{d}}\right)a^2,\quad\mathtt{d}\rightarrow\infty.
\end{align}
\end{subequations}
The same limit (with the slower rate of convergence) is achieved by the Dirichlet box too, as Table~\ref{Table_Dirichlet} demonstrates. Interestingly, all 4D angle-independent states (apart from the lowest Neumann level) possess the same momentum measure $\frac{4}{3}a^2$ that is not influenced by the edge requirement, Eq.~\eqref{FisherCases1e}. It is highly relevant to recall here that for the hydrogen atom its position (momentum) Fisher information decreases (increases) with the dimensionality as well as with the radial index $n$ \cite{Dehesa6,Romera1} what for $I_\rho^{(\mathtt{d})H}$ is consistent with the corresponding behavior of the energy spectrum.
\newpage
%
\begin{sidewaystable}
\caption{Ground-state unitless measures of the Dirichlet dot in terms of its dimensionality $\mathtt{d}$. For comparison, right-hand side of the Shannon uncertainty relation, Eq.~\eqref{ShannonInequality1}, is also shown. The data for $\mathtt{d}=1$ and 2 were taken from the previous researches \cite{Olendski1,Olendski5,Olendski6}.}
\centering 
\begin{tabular}{|c||c|c|c|c||c|c|c||c|c|c|}
\hline 
\multirow{2}{1em}{$\mathtt{d}$}
 &\multicolumn{4}{|c||}{Shannon entropy}&\multicolumn{3}{c||}{Fisher information}&\multicolumn{3}{c|}{Onicescu energy}\\ 
\cline{2-11}
&$\overline{S}_\rho^{(\mathtt{d})D}$&$\overline{S}_\gamma^{(\mathtt{d})D}$&$S_\rho^{(\mathtt{d})D}+S_\gamma^{(\mathtt{d})D}$&$\mathtt{d}(1+\ln\pi)$&$\overline{I}_\rho^{(\mathtt{d})D}$&$\overline{I}_\gamma^{(\mathtt{d})D}$&$I_\rho^{(\mathtt{d})D}I_\gamma^{(\mathtt{d})D}$&$\overline{O}_\rho^{(\mathtt{d})D}$&$\overline{O}_\gamma^{(\mathtt{d})D}$&$O_\rho^{(\mathtt{d})D}O_\gamma^{(\mathtt{d})D}$\\
\hline 
1&-0.30685&2.5189&2.2120&2.1447&39.478&0.13069&5.1595&1.5&0.93366E-1&0.14005\\
2&0.59417&3.8232&4.4174&4.2895&23.133&0.87223&20.177&0.66793&0.29091E-1&0.19431E-1\\
3&0.67558&5.9418&6.6173&6.4342&39.478&1.1307&44.638&0.67207&0.39864E-2&0.26791E-2\\
4&0.65855&8.1545&8.8131&8.5789&58.728&1.3333&78.304&0.74380&0.49433E-3&0.36768E-3\\
5&0.56070&10.445&11.005&10.724&80.763&1.4984&121.02&0.89022&0.56474E-4&0.50274E-4\\
6&0.39418&12.801&13.195&12.868&105.50&1.6367&172.66&1.1389&0.60177E-5&0.68534E-5\\
7&0.16790&15.214&15.382&15.013&132.87&1.7548&233.16&1.5441&0.60352E-6&0.93186E-6\\
8&-0.11132&17.678&17.567&17.158&162.83&1.8574&302.43&2.2040&0.57363E-7&0.12643E-6\\
9&-0.43810&20.188&19.750&19.303&195.32&1.9477&380.43&3.2949&0.51955E-8&0.17119E-7\\
10&-0.80806&22.739&21.931&21.447&230.33&2.0280&467.11&5.1374&0.45041E-9&0.23140E-8\\
20&-6.2927&49.979&43.686&42.895&713.35&2.5296&0.18045E+4&0.24914E+4&0.17688E-20&0.44068E-17\\
30&-14.021&79.396&65.375&64.342&0.14288E+4&2.7891&0.39851E+4&0.10860E+8&0.71574E-33&0.77728E-26\\
40&-23.269&110.30&87.027&85.789&0.23694E+4&2.9540&0.69992E+4&0.21047E+12&0.62188E-46&0.13088E-34\\
50&-33.672&142.33&108.65&107.24&0.35309E+4&3.0704&0.10841E+5&0.12701E+17&0.16800E-59&0.21338E-43\\
60&-45.007&175.27&130.26&128.68&0.49104E+4&3.1580&0.15507E+5&0.19210E+22&0.17672E-73&0.33949E-52\\
70&-57.126&208.98&151.86&150.13&0.65062E+4&3.2269&0.20995E+5&0.62891E+27&0.84244E-88&0.52982E-61\\
80&-69.920&243.36&173.44&171.58&0.83166E+4&3.2829&0.27302E+5&0.40088E+33&0.20303E-102&0.81391E-70\\
90&-83.308&278.32&195.02&193.03&0.10340E+5&3.3294&0.34427E+5&0.45922E+39&0.26869E-117&0.12339E-78\\
100&-97.225&313.81&216.58&214.47&0.12577E+5&3.3688&0.42369E+5&0.88781E+45&0.20831E-132&0.18494E-87\\
200&-257.44&689.43&431.99&428.95&0.46490E+5&3.5818&0.16652E+6&0.71864E+117&0.98579E-293&0.70842E-176\\
300&-443.04&0.10902E+4&647.15&643.42&0.10117E+6&3.6740&0.37169E+6&0.49848E+200&0.36021E-464&0.17956E-264\\
400&-645.26&0.15074E+4&862.19&857.89&0.17642E+6&3.7276&0.65763E+6&0.51265E+290&0.70925E-643&0.36360E-353\\
\hline
\end{tabular}
\label{Table_Dirichlet}
\end{sidewaystable}
%

%
\begin{table}
\caption{The same as in Table~\ref{Table_Dirichlet} but for the Neumann dot. Since the ground-state position Fisher information for this BC is zero, corresponding columns $\overline{I}_\rho^N$ and $I_\rho^NI_\gamma^N$ are not shown}
\centering 
\begin{tabular}{|c||c|c|c|c||c||c|c|c|}
\hline 
\multirow{2}{1em}{$\mathtt{d}$}
 &\multicolumn{4}{|c||}{Shannon entropy}&\multicolumn{1}{c||}{Fisher information}&\multicolumn{3}{c|}{Onicescu energy}\\ 
\cline{2-9}
&$\overline{S}_\rho^{(\mathtt{d})N}$&$\overline{S}_\gamma^{(\mathtt{d})N}$&$S_\rho^{(\mathtt{d})N}+S_\gamma^{(\mathtt{d})N}$&$\mathtt{d}(1+\ln\pi)$&$\overline{I}_\gamma^{(\mathtt{d})N}$&$\overline{O}_\rho^{(\mathtt{d})N}$&$\overline{O}_\gamma^{(\mathtt{d})N}$&$O_\rho^{(\mathtt{d})N}O_\gamma^{(\mathtt{d})N}$\\
\hline 
1&0&2.6834&2.6834&2.1447&0.33333&1&0.10610&0.10610\\
2&1.1447&4.2880&5.4327&4.2895&2&0.31831&0.36575E-1&0.11642E-1\\
3&1.4324&6.7784&8.2108&6.4342&2.4&0.23873&0.54681E-2&0.13054E-2\\
4&1.5963&9.4084&11.005&8.5789&2.6667&0.20264&0.73287E-3&0.14851E-3\\
5&1.6609&12.147&13.808&10.724&2.8571&0.18998&0.89856E-4&0.17071E-4\\
6&1.6424&14.976&16.618&12.868&3&0.19351&0.10219E-4&0.19774E-5\\
7&1.5528&17.880&19.433&15.013&3.1111&0.21165&0.10888E-5&0.23044E-6\\
8&1.4009&20.850&22.250&17.158&3.2&0.24638&0.10952E-6&0.26984E-7\\
9&1.1935&23.877&25.071&19.303&3.2727&0.30317&0.10464E-7&0.31723E-8\\
10&0.93616&26.957&27.894&21.447&3.3333&0.39213&0.95420E-9&0.37417E-9\\
20&-3.6571&59.836&56.178&42.895&3.6364&38.749&0.56057E-20&0.21722E-18\\
30&-10.728&95.237&84.508&64.342&3.75&0.45630E+5&0.30489E-32&0.13912E-27\\
40&-19.441&132.30&112.86&85.789&3.8095&0.27741E+9&0.33614E-45&0.93250E-37\\
50&-29.385&170.60&141.21&107.24&3.8462&0.57796E+13&0.11104E-58&0.64179E-46\\
60&-40.316&209.89&169.57&128.68&3.8710&0.32297E+18&0.13916E-72&0.44944E-55\\
70&-52.071&250.01&197.94&150.13&3.8889&0.41114E+23&0.77500E-87&0.31863E-64\\
80&-64.531&290.84&226.30&171.58&3.9024&0.10608E+29&0.21490E-101&0.22797E-73\\
90&-77.611&332.28&254.67&193.03&3.9130&0.50824E+34&0.32321E-116&0.16427E-82\\
100&-91.241&374.28&283.04&214.47&3.9216&0.42226E+40&0.2819E0-131&0.11903E-91\\
200&-249.27&816.05&566.78&428.95&3.9604&0.17989E+109&0.31236E-291&0.56192E-183\\
300&-433.31&0.12839E+4&850.54&643.42&3.9735&0.15290E+189&0.19891E-462&0.30413E-274\\
400&-634.29&0.17686E+4&0.11343E+4&857.89&3.9801&0.29303E+276&0.59473E-641&0.17427E-365\\
\hline
\end{tabular}
\label{Table_Neumann}
\end{table}

Momentum Onicescu energy $\overline{O}_\gamma^{(\mathtt{d})}$ for either BC monotonically decreases with the dimensionality whereas the  Dirichlet (Neumann) position functional reaches minimum at $\mathtt{d}=3$ ($\mathtt{d}=5$). Since the Onicescu energy quantitatively describes a deviation from the most probable distribution and since the momentum waveform $\overline{{\cal K}}_\gamma^{(\mathtt{d})}(k)$ flattens with the growth of the dimensionality becoming more and more uniform, as exemplified by the corresponding panel of Fig.~\ref{Fig_Waveforms}, the disequilibrium decreases. On the other hand, left upper subplot manifests a buildup of the position function at the origin with $\mathtt{d}$ growing. This sharpening of the function $\overline{{\cal R}}_\rho^{(\mathtt{d})}(r)$  destroys the homogeneity increasing in this way the corresponding measure. The product of the two functionals $O_\rho^{(\mathtt{d})}O_\gamma^{(\mathtt{d})}$ decreases for both BCs.

\section{R\'{e}nyi and Tsallis entropies}\label{sec_Renyi}
\begin{figure}
\centering
\includegraphics[width=\columnwidth]{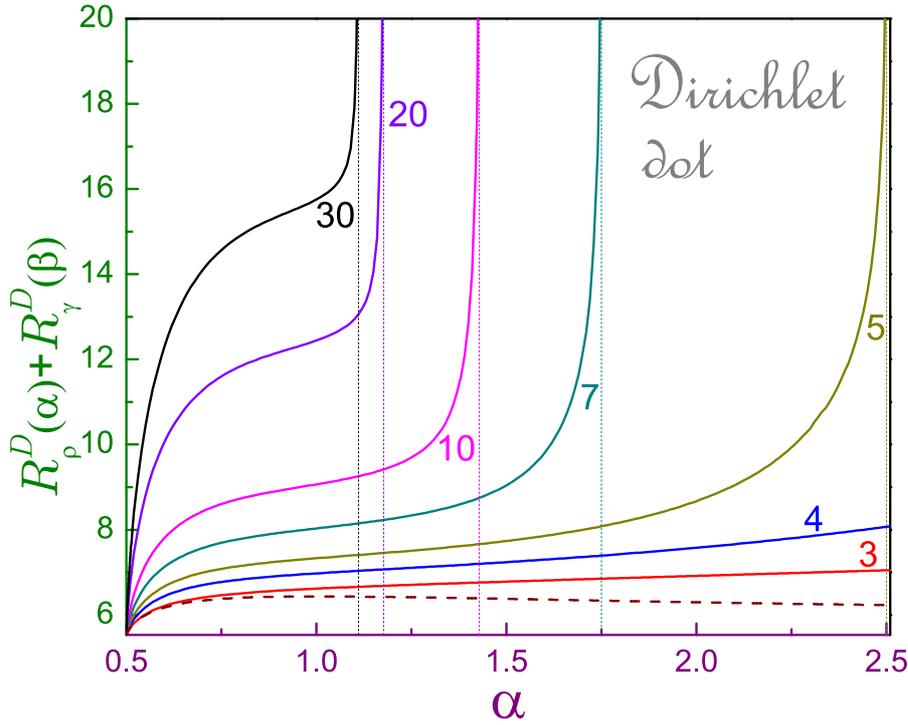}
\caption{\label{Fig_RenyiUncertaintyDirichlet}
Dirichlet ground-state R\'{e}nyi uncertainty relation, Eq.~\eqref{RenyiUncertainty1}, in terms of the coefficient $\alpha$ with the conjugation parameter $\beta$ given by Eq.~\eqref{Beta1}, as it follows from Eq.~\eqref{RenyiUncertainty2}. Numbers at the curves denote the dimensionality $\mathtt{d}$. Thin dotted vertical lines are the corresponding thresholds from Eq.~\eqref{Ddimensional2_D}. Dashed line depicts right-hand side of the R\'{e}nyi uncertainty relation from Eq.~\eqref{RenyiUncertainty1} at $\mathtt{d}=3$. For comparison, the sums $R_\rho(\alpha)+R_\gamma(\beta)$ are offset by $(\mathtt{d}-3)\ln(2\pi)$.}
\end{figure}

Since at the zero parameter $\alpha$ the integrand in Eq.~\eqref{Renyi1_R} degenerates to unity, the position R\'{e}nyi entropy transforms, independently of the edge condition and the orbital, into the logarithm of the volume of the hyperball \cite{Olver1}:
\begin{equation}\label{RenyiPosition0}
R_{\rho_{n,l,\left\{\mu\right\}}}^{(\mathtt{d})}(0)=\ln\frac{\pi^{\mathtt{d}/2}a^\mathtt{d}}{\Gamma\left(\frac{\mathtt{d}}{2}+1\right)}.
\end{equation}
Note that for the hydrogen atom with the semi-infinite radial range of integration, the corresponding functional diverges at $\alpha$ approaching zero, see Eq.~\eqref{HydrogenRenyi1_Position} below. In the opposite limit of the extremely huge R\'{e}nyi parameters, one uses the relation:
\begin{equation}\label{RenyiInfinite1}
R_{\rho,\gamma}(\infty)=-\ln\!\left(\!\begin{array}{c}
\rho_{max}\\\gamma_{max}
\end{array}\!\right);
\end{equation}
for example, for the spherically symmetric orbitals, $l=0$, the global maxima of both position, ${\cal R}_{n,0}^{(\mathtt{d})}(r)$, as well as momentum, ${\cal K}_{n,0}^{(\mathtt{d})}(k)$, waveforms are achieved for the Dirichlet BC at the zero radius and momentum, respectively, Eqs.~\eqref{DirichletPosition1} and \eqref{MomentumFunction1D}, and then:
\begin{subequations}\label{RenyiInfinite2}
\begin{align}\label{RenyiInfinite2Position}
R_{\rho_{n,0,\left\{0\right\}}}^{(\mathtt{d})D}(\infty)&=\mathtt{d}\ln a-2\ln\frac{{\cal Y}_{0,\left\{0\right\}}^{(\mathtt{d})}\left(\Omega_{\mathtt{d}-1}\right)}{\mathtt{d}_1(\mathtt{d}-4)!!\left|j_1^{(\mathtt{d})}\left(j_{\mathtt{d}_1,n}\right)\right|}\\
\label{RenyiInfinite2Momentum}
R_{\gamma_{n,0,\left\{0\right\}}}^{(\mathtt{d})D}(\infty)&=-\mathtt{d}\ln a-2\ln\frac{2^{(1-\mathtt{d})/2}(d-2){\cal Y}_{0,\left\{0\right\}}^{(\mathtt{d})}\left(\Omega_{\mathtt{d}-1}\right)}{j_{\mathtt{d}_1,n}\mathtt{d}_1\Gamma(\mathtt{d}/2)},
\end{align}
\end{subequations}
with ${\cal Y}_{0,\left\{0\right\}}^{(\mathtt{d})}\left(\Omega_{\mathtt{d}-1}\right)$ defined above, Eq.~\eqref{Angular1}. Since the R\'{e}nyi entropy is a monotonically decreasing function of its parameter, Eqs.~\eqref{RenyiPosition0} and \eqref{RenyiInfinite2Position} determine the range inside which its Dirichlet position component varies as the coefficient $\alpha$ scans the positive axis. Considering momentum R\'{e}nyi entropy, its dimensionless part for, e.g., the Dirichlet dot contains the logarithm of the following improper integral:
\begin{equation}\label{Integral1}
\int_0^\infty\!\!z^{\mathtt{d}-1}\!\left|\frac{j_{l+\mathtt{d}_1,n}}{j_{l+\mathtt{d}_1,n}^2-z^2}j_l^{(\mathtt{d})}(z)\right|^{2\alpha}dz.
\end{equation}
Applying to it a comparison convergence test \cite{Fikhtengolts1}, one arrives at the conclusion that the momentum R\'{e}nyi and Tsallis measures do exist not at any positive parameter but only at the coefficient $\alpha$ greater than the lower threshold given by Eq.~\eqref{Ddimensional1_D}. In the same way, one proves a correctness of the Neumann threshold, Eq.~\eqref{Ddimensional1_N}. And then, the upper thresholds of the R\'{e}nyi uncertainty relation, Eqs.~\eqref{Ddimenional3}, follow straightforwardly. 

\begin{figure}
\centering
\includegraphics[width=\columnwidth]{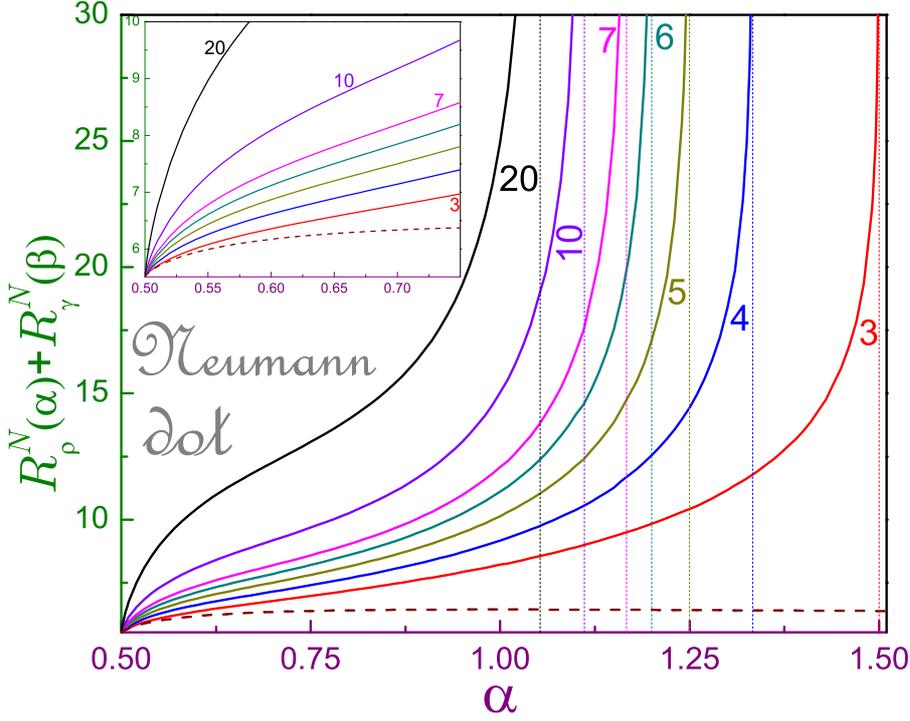}
\caption{\label{Fig_RenyiUncertaintyNeumann}
The same as in Fig.~\ref{Fig_RenyiUncertaintyDirichlet} but for the Neumann BC. Thin dotted vertical lines are the corresponding thresholds from Eq.~\eqref{Ddimensional2_N}. Note different ranges of the axes as compared to Fig.~\ref{Fig_RenyiUncertaintyDirichlet}. Inset shows the behavior near $\alpha=\frac{1}{2}$.}
\end{figure}

\begin{figure}
\centering
\includegraphics[width=\columnwidth]{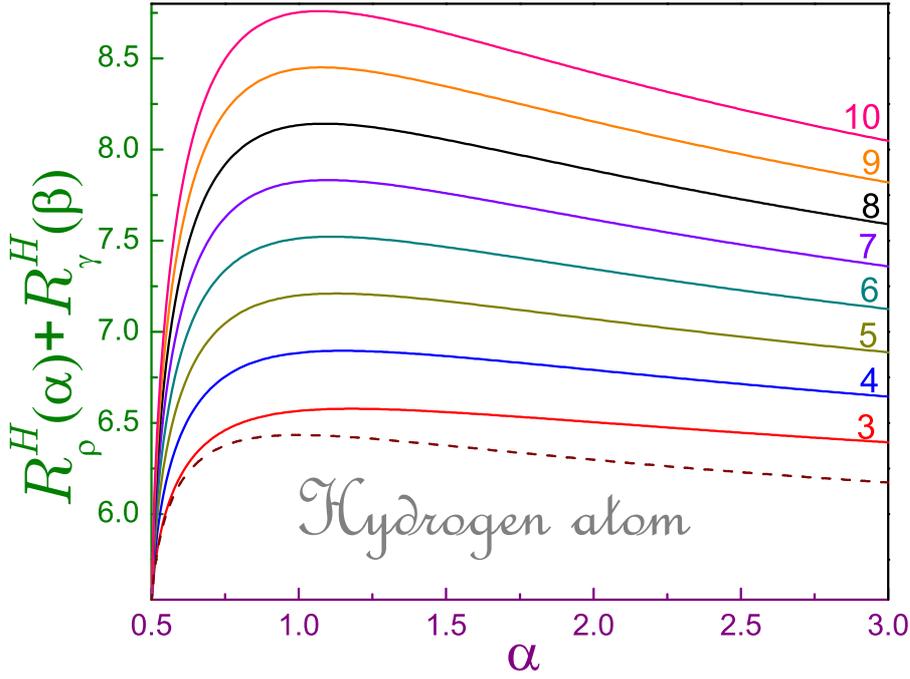}
\caption{\label{Fig_RenyiUncertaintyHydrogen}
The same as in Fig.~\ref{Fig_RenyiUncertaintyDirichlet} but for the hydrogen-like ion. Since the sum $R_\rho(\alpha)+R_\gamma(\beta)$ for this geometry is defined at arbitrary $\alpha$ greater than one half, no vertical lines indicating thresholds are drawn.}
\end{figure}

Figs.~\ref{Fig_RenyiUncertaintyDirichlet} and \ref{Fig_RenyiUncertaintyNeumann} depict, at several dimensionalities, the ground-state R\'{e}nyi uncertainty relation, Eq.~\eqref{RenyiUncertainty1}, for the Dirichlet and Neumann dots, respectively. For the 3D structure, as it follows from Eq.~\eqref{Ddimensional2_D}, the former edge requirement defines this inequality at any positive $\alpha$ greater than one half but the Neumann BC limits this interval from above by the upper threshold of 3/2, Eq.~\eqref{Ddimensional2_N}, as shown in Fig.~\ref{Fig_RenyiUncertaintyNeumann}. For $\mathtt{d}=4$, the Dirichlet sum diverges at $\alpha=4$ (not shown in Fig.~\ref{Fig_RenyiUncertaintyDirichlet}), Eq.~\eqref{Ddimensional2_D}, whereas for the Neumann geometry it takes place at $4/3$, as Eq.~\eqref{Ddimensional2_N} implies and Fig.~\ref{Fig_RenyiUncertaintyNeumann} exemplifies. The increase of the R\'{e}nyi sum with the increase of the parameter $\alpha$ means a loss of the information about the corresponding state and at $\alpha$ approaching $\alpha_R(\mathtt{d})$ we know nothing about its momentum. As is seen from the figures, the speed of change $\frac{d}{d\alpha}\left[R_\rho^{(\mathtt{d})}(\alpha)+R_\gamma^{(\mathtt{d})}(\beta)\right]$ increases with the dimensionality and, of course, inequality~\eqref{RenyiUncertainty1} is always satisfied. Let us note that for the hydrogen atom the threshold \cite{Aptekarev2}
\begin{equation}\label{DdimensionalHydrogen1}
\alpha_{TH}^H=\frac{1}{2}\frac{\mathtt{d}}{\mathtt{d}+l+1}
\end{equation}
is always smaller than one half what means that the corresponding uncertainty relation, independently of the dimensionality, is defined at any parameter $\alpha$ greater than $1/2$. It is worthwhile also to remark here that the hyperdot limits, Eqs.~\eqref{Ddimensional1}, are level independent ones whereas their hydrogen counterpart from Eq.~\eqref{DdimensionalHydrogen1} is influenced, in addition to $\mathtt{d}$, by the orbital quantum number $l$; in particular, extremely huge indices subdue it to zero. Hydrogen ground-state R\'{e}nyi entropies are calculated analytically: with the help of Eqs.~\eqref{Hydrogen1} where one puts $n=1$ and $l=0$, the result is:
\begin{subequations}\label{HydrogenRenyi1}
\begin{align}\label{HydrogenRenyi1_Position}
R_{\rho_{1,0,\left\{0\right\}}}^{(\mathtt{d})H}(\alpha)&=\mathtt{d}\ln(\lambda r_0)-\ln{{\cal Y}_{0,\left\{0\right\}}^{(\mathtt{d})}}^{\!\!\!\!\!2}\left(\Omega_{\mathtt{d}-1}\right)+\ln(\mathtt{d}-1)!-\frac{\mathtt{d}}{1-\alpha}\ln\alpha+\frac{\alpha}{1-\alpha}\ln\frac{\mathtt{d}-1}{4\lambda}\\
R_{\gamma_{1,0,\left\{0\right\}}}^{(\mathtt{d})H}(\alpha)&=-\mathtt{d}\ln(\lambda r_0)-\ln{{\cal Y}_{0,\left\{0\right\}}^{(\mathtt{d})}}^{\!\!\!\!\!2}\left(\Omega_{\mathtt{d}-1}\right)+\frac{3\mathtt{d}\alpha-\mathtt{d}-1}{1-\alpha}\ln2\nonumber\\
\label{HydrogenRenyi1_Momentum}
+&\frac{\alpha}{1-\alpha}\ln\frac{\lambda\Gamma^2\left(\frac{\mathtt{d}-1}{2}\right)}{\pi(\mathtt{d}-2)!}+\frac{1}{1-\alpha}\ln\frac{\Gamma\left(\alpha(\mathtt{d}+1)-\frac{\mathtt{d}}{2}\right)\Gamma\left(\frac{\mathtt{d}}{2}\right)}{\Gamma(\alpha(\mathtt{d}+1))}.
\end{align}
\end{subequations}
As stated at the beginning of this Section, the position measure diverges at the vanishing coefficient, as Eq.~\eqref{HydrogenRenyi1_Position} vividly demonstrates. Next, Eq.~\eqref{HydrogenRenyi1_Momentum} confirms the thresholds from Eq.~\eqref{DdimensionalHydrogen1}. Using Eqs.~\eqref{HydrogenRenyi1}, one can write and analyze the corresponding uncertainty relation, Eq.~\eqref{RenyiUncertainty1}. Fig.~\ref{Fig_RenyiUncertaintyHydrogen} shows its left-hand side in terms of the coefficient $\alpha$ for several $\mathtt{d}$. Striking differences with the hyperballs are seen; first, the range of the R\'{e}nyi factor where the sum is defined stretches to infinity. Second, the function $R_{\rho_{1,0,\left\{0\right\}}}^{(\mathtt{d})H}(\alpha)+R_{\gamma_{1,0,\left\{0\right\}}}^{(\mathtt{d})H}\!\left(\frac{\alpha}{2\alpha-1}\right)$ is not a monotonic one: at some dimensionality-dependent $\alpha_{max}^{(\mathtt{d})H}$ it reaches a maximum after which smoothly decreases. With the help of Eqs.~\eqref{HydrogenRenyi1}, it is easy to derive its leading term at the unrestrictedly increasing R\'{e}nyi coefficient:
\begin{equation}\label{Asymptote1}
R_{\rho_{1,0,\left\{0\right\}}}^{(\mathtt{d})H}(\alpha)+R_{\gamma_{1,0,\left\{0\right\}}}^{(\mathtt{d})H}\!\!\left(\frac{\alpha}{2\alpha-1}\right)\rightarrow-2\ln{{\cal Y}_{0,\left\{0\right\}}^{(\mathtt{d})}}^{\!\!\!\!\!2}\left(\Omega_{\mathtt{d}-1}\right)+\ln\frac{\pi^{1/2}(\mathtt{d}-1)!}{2}-\ln\frac{\Gamma\left(\frac{\mathtt{d}+1}{2}\right)}{\Gamma\left(\frac{\mathtt{d}}{2}\right)},\,\alpha\rightarrow\infty.
\end{equation}
To find a location of the extremum $\alpha_{max}^{(\mathtt{d})H}$, one needs to take a derivative of the sum and equate it to zero. However, the obtained equation that, in addition to the $\Gamma$-functions of the miscellaneous arguments, contains also Digamma functions, can not be solved analytically. Numerically, it is found that $\alpha_{max}^{(\mathtt{3})H}=1.1798\ldots$, $\alpha_{max}^{(\mathtt{4})H}=1.1498\ldots$, $\alpha_{max}^{(\mathtt{5})H}=1.1272\ldots$, etc. As the dimensionality unboundedly grows, the position of the maximum asymptotically approaches the Shannon case:
\begin{equation}\label{Asymptote2}
\alpha_{max}^{(\mathtt{d})H}\rightarrow1,\quad\mathtt{d}\rightarrow\infty.
\end{equation}

Another important property one needs to examine is the fact that both R\'{e}nyi, Eq.~\eqref{RenyiUncertainty1}, and Tsallis, Eq.~\eqref{TsallisInequality1}, inequalities for the ground state turn into the identities at $\alpha=1/2$. Explanation of this phenomenon for the hyperdots is similar to the previously discussed geometries \cite{Olendski1,Olendski2,Olendski3,Olendski4}. Consider, for example, the R\'{e}nyi relation. At $\alpha=1/2$ the first item in its left-hand side reads:
$$2\ln\int_{\mathcal{D}_\rho^{(\mathtt{d})}}\left|\Psi_\mathtt{n}^{(\mathtt{d})}({\bf r})\right|d{\bf r}.$$ Absolute value of the function is equal to the function itself when, first, the function is real what, due to Eqs.~\eqref{Separation1} and \eqref{AngularFunction1}, means that $m=0$, and, second, when it does not change its sign, what, according to the same equations, takes place for the ground state only, $n=1$, $l=0$. Obviously, the first requirement is absorbed by the second one. Then, $$R_{\rho_{1,0,\left\{0\right\}}}^{(\mathtt{d})}\left(\frac{1}{2}\right)=2\ln\int_{\mathcal{D}_\rho^{(\mathtt{d})}}\Psi_{1,0,\left\{0\right\}}^{(\mathtt{d})}({\bf r})d{\bf r}.$$Recalling the Fourier transform, Eq.~\eqref{Fourier1_1}, the above turns to
$$R_{\rho_{1,0,\left\{0\right\}}}^{(\mathtt{d})}\left(\frac{1}{2}\right)=\mathtt{d}\ln(2\pi)+2\ln\Phi_{1,0,\left\{0\right\}}^{(\mathtt{d})}({\bf 0}).$$ According to Eq.~\eqref{RenyiInfinite1}, the momentum item in the same limit, i.e., at $\beta=\infty$, becomes $-2\ln\left|\Phi_\mathtt{n}^{(\mathtt{d})}({\bf k})\right|_{max}$, and since, as discussed above, the global maximum of the ground-state function is achieved at the zero momentum, the left-hand side of the R\'{e}nyi relation, Eq.~\eqref{RenyiUncertainty1}, simplifies to $\mathtt{d}\ln(2\pi)\approx1.8379\mathtt{d}$. But this is exactly what the right-hand side turns to at $\alpha\rightarrow\frac{1}{2}$ \cite{Olendski3}:
$$-\frac{\mathtt{d}}{2}\left(\frac{1}{1-\alpha}\ln\frac{\alpha}{\pi}+\frac{1}{1-\beta}\ln\frac{\beta}{\pi}\right)\rightarrow\mathtt{d}(\ln2\pi-[1+\ln(2\alpha-1)](2\alpha-1)+\ldots),\,\alpha\rightarrow\frac{1}{2}.$$Transformation at $\alpha=1/2$ of the ground-state inequality into the identity at any dimensionality and BC is clearly seen in Figs.~\ref{Fig_RenyiUncertaintyDirichlet} - \ref{Fig_RenyiUncertaintyHydrogen} with the magnified picture presented in the inset of Fig.~\ref{Fig_RenyiUncertaintyNeumann}. For any other level, relation~\eqref{RenyiUncertainty1} at any available parameter takes the form of a strict inequality.

\begin{figure}
\centering
\includegraphics[width=\columnwidth]{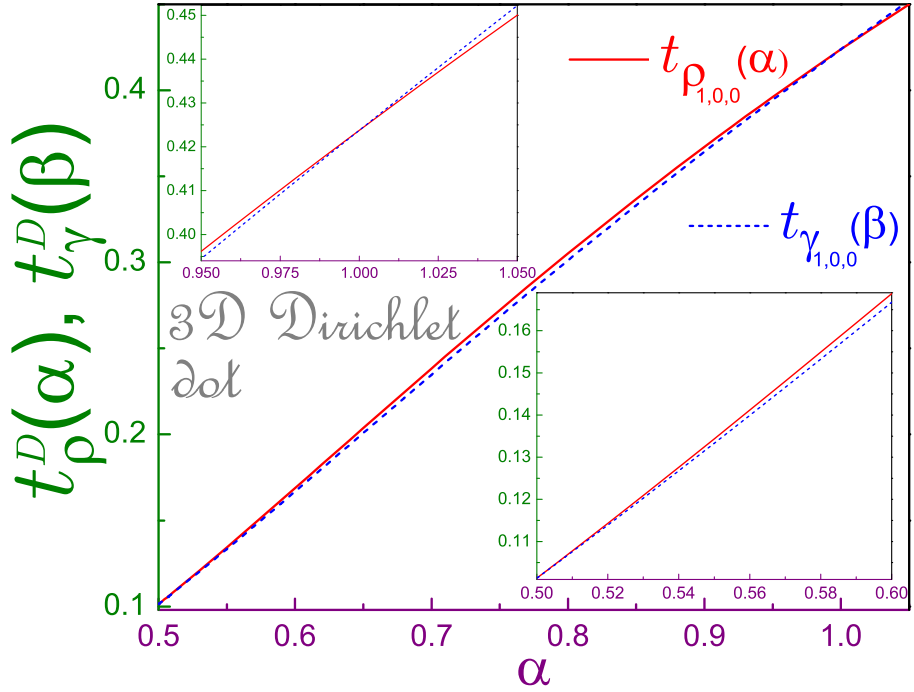}
\caption{\label{Fig_Tsallis}
Dimensionless Tsallis position $t_{\rho_{1,0,0}}^{\mathtt{3}D}(\alpha)$, Eq.~\eqref{TsallisInequality2_R}, solid line, and momentum $t_{\gamma_{1,0,0}}^{\mathtt{3}D}(\beta)$, Eq.~\eqref{TsallisInequality2_K}, (dotted curve) functions in terms of parameter $\alpha$ for the $\mathtt{3}$D Dirichlet ground-state orbital. Lower right inset exhibits functions close to the left border $\alpha=1/2$, and upper left subplot shows $t_{\rho_{1,0,0}}^{\mathtt{3}D}(\alpha)$ and $t_{\gamma_{1,0,0}}^{\mathtt{3}D}(\beta)$ in $\alpha=1$ vicinity.}
\end{figure}

As the integral of the form from Eq.~\eqref{Integral1} enters into the expression of the Tsallis measure, its momentum counterpart does exist also only when its parameter is greater than the thresholds from Eqs.~\eqref{Ddimensional1}. But since the associated uncertainty relation, Eq.~\eqref{TsallisInequality1}, is valid inside the interval with the upper edge equal to unity, Eq.~\eqref{Sobolev2}, the R\'{e}nyi constraint, Eq.~\eqref{Ddimensional2}, does not influence it. At $\alpha=1$, Eqs.~\eqref{TsallisInequality1} and \eqref{Sobolev1} for any bound state become identities with their either side being $\pi^{-\mathtt{d}/4}$. In addition, using the arguments provided above for the R\'{e}nyi entropies, one sees that the ground orbital transforms these relations into the equality at $\alpha=1/2$ too when their either side contains $\Phi_{1,0,\left\{0\right\}}^{(\mathtt{d})}({\bf 0})$. Since both just mentioned quantities decrease with the dimensionality, Fig.~\ref{Fig_Tsallis} presents dimensionless position $t_{\rho_{1,0,0}}^{(\mathtt{3})D}(\alpha)$ and momentum $t_{\gamma_{1,0,0}}^{(\mathtt{3})D}(\beta)$ components of the $\mathtt{3}$D Dirichlet Tsallis relation
\begin{subequations}\label{TsallisInequality2}
\begin{align}\label{TsallisInequality2_R}
t_{\rho_{n,l,\left\{\mu\right\}}}^{(\mathtt{d})D}(\alpha)&=t_{ang_{l,\left\{\mu\right\}}}^{(\mathtt{d})}(\alpha)\left(\frac{\alpha}{\pi}\right)^{\mathtt{d}/(4\alpha)}\frac{2^{1/2}}{\left|j_{l+1}^{(\mathtt{d})}(j_{l+\mathtt{d}_1,n})\right|}\left(\int_0^1\left|j_l^{(\mathtt{d})}\left(j_{l+\mathtt{d}_1,n}z\right)\right|^{2\alpha}z^{\mathtt{d}-1}dz\right)^{\!\!1/(2\alpha)}\\
\label{TsallisInequality2_K}
t_{\gamma_{n,l,\left\{\mu\right\}}}^{(\mathtt{d})D}(\beta)&=t_{ang_{l,\left\{\mu\right\}}}^{(\mathtt{d})}(\beta)\left(\frac{\beta}{\pi}\right)^{\mathtt{d}/(4\beta)}\!\!2^{(3-\mathtt{d})/2}\frac{(\mathtt{d}-2)!!}{\Gamma(\mathtt{d}/2)}j_{l+\mathtt{d}_1,n}\!\left(\int_0^\infty\left|\frac{j_l^{(\mathtt{d})}(z)}{j_{l+\mathtt{d}_1,n}^2-z^2}\right|^{2\beta}\!z^{\mathtt{d}-1}dz\right)^{\!\!1/(2\beta)}
\intertext{with the angular part $t_{ang_{l,\left\{\mu\right\}}}^{(\mathtt{d})}(\omega)$ at any BC being}
\label{TsallisInequality2_Ang}
t_{ang_{l,\left\{\mu\right\}}}^{(\mathtt{d})}(\omega)&=\left(\int\left|{{\cal Y}_{l,\left\{\mu\right\}}^{(\mathtt{d})}}(\Omega_{\mathtt{d}-1})\right|^{2\omega}d\Omega_{\mathtt{d}-1}\right)^{1/(2\omega)}.
\end{align}
\end{subequations}
It is seen that inequality~\eqref{TsallisInequality1}, as it should be, holds true at the Tsallis coefficient $\alpha$ lying between one half and unity, Eq.~\eqref{Sobolev2}, at the right edge of this interval both sides of it degenerate to dimensionless $\pi^{-3/4}=0.42377\ldots$ whereas at $\alpha=1/2$ they are equal again with the value of $\Phi_{1,0,0}^{(\mathtt{3})D}({\bf 0})$ being $a^{3/2}\pi^{-2}=a^{3/2}0.10132\ldots$.

\section{Concluding remarks}\label{sec_Conclusions}
Each spatially confined system one way or another interacts with the exterior environment. To elucidate the character and intensity of this interaction, theoretical physics supplements the wave equation describing the fields inside the system by the BC that the corresponding solution has to satisfy at the surface between the object and surroundings. Linear relation linking together the position waveform and its normal derivative at the hypersurface takes the form of the Robin BC \cite{Gustafson1}:
\begin{equation}\label{Robin1}
\left.{\bf n}{\bm\nabla}\Psi(\bf r)\right|_{\cal S}=\frac{1}{\Lambda}\Psi(\bf r)|_{\cal S},
\end{equation}
where the extrapolation length $\Lambda$ in general is a function of the position on the interface and can take complex values. In the present research, its two limiting cases $\Lambda=0$ (Dirichlet requirement) and $\Lambda=\infty$ (Neumann demand) have been compared with the emphasis on the quantum information measures of the $\mathtt{d}$D spherical hyperball. Regarding the Shannon entropy, it has been discovered that independently of the dimensionality the geometry with latter BC provides less total information about the position and motion than the Dirichlet configuration. Among many other findings, one needs to mention that the type of the edge requirement strongly affects the regions where the momentum R\'{e}nyi/Tsallis functionals are defined, Eqs.~\eqref{Ddimensional1}, and where the R\'{e}nyi uncertainty relation is valid, Eqs.~\eqref{Ddimensional2}.

Discussed above measures can be used as building blocks for defining miscellaneous complexities: $e^SO$ \cite{Catalan1}, $\frac{1}{2\pi e}e^{2S/\mathtt{d}}I$ \cite{Dembo1,Vignat1}, $e^{R(\alpha)}O$ \cite{Antolin1,Nath1} and others, see Refs.~\cite{Sen1,Toranzo6}. Some of them have been analyzed for the different $\mathtt{d}$D geometries with $\mathtt{d}\geq3$ \cite{SobrinoColl1,Dehesa7,Sen1,Toranzo6,LopezRosa2}. Based on the results provided in the present research, one can expand this knowledge; for example, it is known that for any $\mathtt{d}$D space neither position nor momentum component of the first product can never be less than unity, $e^SO\geq1$ \cite{LopezRosa3}, with the equality being reached only for the uniform distribution with a finite volume support; indeed, above relation is tightened by the position component of the lowest Neumann orbital, as it immediately follows from Eqs.~\eqref{NeumannPositionMeasures1_Shannon} and \eqref{NeumannPositionMeasures1_Onicescu}. Picking up corresponding Dirichlet, Table~\ref{Table_Dirichlet}, entries, one sees that for this BC the equation above in this paragraph takes the form of a strict inequality independently of the dimensionality.

In the present research, the dimensionality of the dot was kept \textit{integer}. An interesting extension useful in quantum-field theories of the $\mathtt{d}$D Euclidean space is a geometry of the \textit{fractional} dimensionality when the Dirichlet energies below some negative orbital-dependent $\mathtt{d}$ become complex \cite{Bender1,Bender2}. Let us also note that preliminary results indicate that the thresholds from Eqs.~\eqref{Ddimensional1_N} and \eqref{Ddimensional2_N} remain valid for any finite positive or negative Robin length $\Lambda$ from Eq.~\eqref{Robin1} what singles out the Dirichlet BC, $\Lambda=0$, as the only one having the unique verges, Eqs.~\eqref{Ddimensional1_D} and \eqref{Ddimensional2_D}; however, a detailed treatment of the quantum-information measures of this configuration requires a separate careful investigation.

\begin{acknowledgements}
Research was supported by Competitive Research Project No. 2002143087 from the Research Funding Department, Vice Chancellor for Research and Graduate Studies, University of Sharjah.
\end{acknowledgements}

%
\section*{Conflict of interest}
The author  declares that he has no conflict of interest.

\end{document}